\documentclass[reprint, secnumarabic,amssymb, nobibnotes, aps, prx]{revtex4-1}

\usepackage{color}
\usepackage{epsfig, graphics, graphicx, subfigure, wrapfig, lipsum, longtable,array}
\usepackage{verbatim, import}
\usepackage{dcolumn}
\usepackage{bm}
\usepackage{amsmath, braket}
\usepackage{array}
\usepackage{xr}

\newcolumntype{X}[1]{>{\centering\arraybackslash\hspace{0pt}}p{#1}}
\newcolumntype{M}[1]{ >{\centering\arraybackslash}m{#1}}
\newcommand{\roml}[1]{\lowercase\expandafter{\romannumeral #1\relax}}
\newcommand{\romu}[1]{\uppercase\expandafter{\romannumeral #1\relax}}

\begin{document}

\title{High-throughput computational discovery of 40 ultralow thermal conductivity and 20 highly anisotropic crystalline materials}

\author{Ankit Jain}
\email{a\_jain@iitb.ac.in}
\author{Harish P Veeravenkata}
\author{Shravan Godse}
\author{Yagyank Srivastava}
\affiliation{Mechanical Engineering Department, IIT Bombay, India}
\date{\today}%

\begin{abstract}
We performed ab-initio driven density functional theory-based high throughput computations to search for materials with low thermal conductivity and high thermal transport anisotropy. We shortlisted a pool of 429 stable ternary semiconductors from the Materials Project and obtained phonon thermal conductivity by solving the Boltzmann transport equation on 225 materials.  We found the lowest thermal conductivity of $0.16$ $\text{W/m-K}$ in $\text{SbRbK}_2$ and 40 materials with a thermal conductivity lower than $1$ $\text{W/m-K}$ at 300 K. For anisotropic thermal transport, we have identified six materials with anisotropy larger than 5 and 20 with thermal transport anisotropy higher than the largest reported literature value. 
\end{abstract}
\maketitle

Thermal conductivity is an important material property that plays a critical role in determining the performance and efficiency of devices in various applications such as thermoelectric energy generation, thermal insulation, and memory storage \cite{clarke2003,dames2005,lindsay2018}. For many of these applications, low thermal conductivity semiconducting solids are desired \cite{clarke2003,dames2005,lindsay2018} in which thermal transport is due to atomic vibrations, i.e., phonons \cite{ziman1960}. Traditionally, the search for noble low-thermal conductivity materials is led by experiments based on a trial-error approach guided by simple physics principles such as the presence of heavy atomic species and complex material unitcell \cite{slack1973}. From the past decade, ab-initio driven density functional theory (DFT) based calculations are also proving promising to search for noble materials with desired thermal properties \cite{esfarjani2008, esfarjani2011, lindsay2018, mcgaughey2019}. Such calculations are already used to understand the thermal transport physics in many materials and excellent agreement  with experiments (wherever possible) is observed in predicted thermal conductivities \cite{esfarjani2011, lindsay2013b, lindsay2018, jain2020}. The notable success case for such calculation is that of boron arsenide, for which computations predicted ultrahigh thermal conductivity\cite{lindsay2013a} (only next to that of the diamond), which was later verified in experiments \cite{li2018, kang2018, tian2018}. 

Though such DFT driven thermal conductivity calculations are instrumental in correctly describing the material thermal transport physics and thus accelerating noble material discovery compared to the experimental trial-error approach by manifold, these calculations are computationally expensive and require, for instance, a single computer running for 10,000-100,000 hours to arrive at a thermal conductivity of a single material \cite{esfarjani2008}. Thus, the application of such calculations is limited to simple material systems, and computational exploration of new materials is restricted to the simple substitution of one or two atomic species in known material systems \cite{xia2020b,miyazaki2021,  he2021, pal2021}. 

In the past few years, there have been significant developments to address these challenges, especially related to the extraction of anharmonic force constants which are otherwise responsible for more than 80-90\% computational cost of these calculations \cite{esfarjani2008, esfarjani2011}. These developments, such as the stochastic thermal snapshot technique to extract the temperature-dependent anharmonic force constants, now allow predicting the thermal conductivity of materials at 1-3 orders of magnitude reduced computational cost \cite{west2006, hellman2013, shulumba2017, ravichandran2018}.

Making use of such developments, in this work, we performed true high-throughput calculations to search for noble material systems with low phonon thermal conductivity and high thermal transport anisotropy. Unlike past studies where the search is restricted to a certain class of materials \cite{ xia2020b,miyazaki2021,  he2021, pal2021}, we have performed full DFT-driven calculations on more than 230 ternary compounds spanning 43 chemical species across 32 spacegroups. Since phonon thermal conductivity follows simple $1/T$ dependence, where $T$ is the temperature \cite{jain2014}, all calculations are performed only at 300 K. All calculations are performed systematically using the same set of well-converged simulation parameters by ensuring thermodynamic and structural stability (no imaginary phonon modes) of all compounds. Our major findings from this high-throughput search are:
\begin{enumerate}
    \item The lowest thermal conductivity obtained is $0.16$ W/m-K at 300 K. Further, 40 materials are identified to have thermal conductivity lower than 1 W/m-K. 
    \item The obtained highest thermal transport anisotropy is $12.6$ amongst covalently bonded compounds. Further, six materials are identified to have thermal transport anisotropy higher than 5 and 20 materials are identified to have anisotropy larger than the highest literature reported value.
\end{enumerate}

\section*{Results}
\label{sec_screening}
\begin{figure*}
\begin{center}
\epsfbox{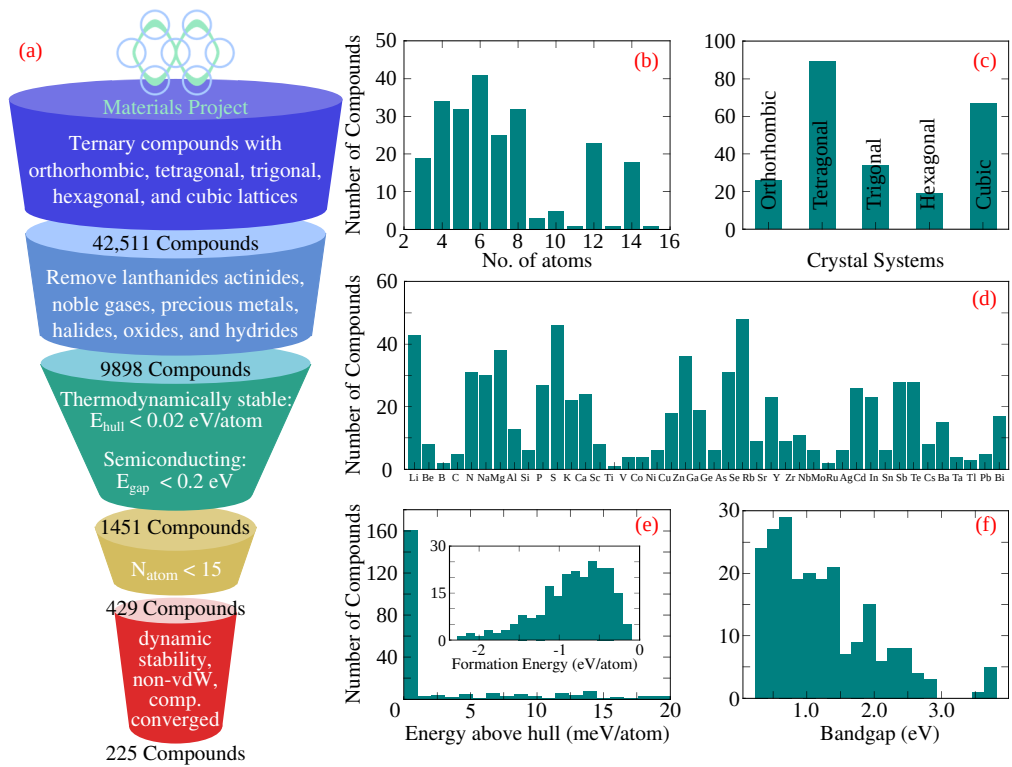}
\end{center}
\caption{(a) The employed filtering criterion to select a pool of 429 stable ternary semiconductors from the Materials Project \cite{materialsProject}. The diversity in terms of (b) number of atoms in the primitive unit-cell, (c) crystalline system, (d) participating atomic species, (e) stability, and (f) electronic bandgap of final 225 materials for which the Boltzmann transport equation is solved. The energies above the convex hull are with respect to all materials available on the Material Project. The energies above the convex hull, the formation energies, and the electronic bandgaps are taken from the Materials Project \cite{materialsProject}.}
\label{fig_screening}
\end{figure*}

Our material screening process for selecting compounds on which thermal conductivity calculations are carried out is detailed in Fig.~\ref{fig_screening}(a). We start  by first selecting all ternary compounds from the Materials Project \cite{materialsProject} within the orthorhombic, tetragonal, trigonal, hexagonal, and cubic spacegroups resulting in a total of 42,511 compounds. Of these, we removed compounds containing lanthanides and actinides, noble gases, and precious metals (Au, Pt, Pd, Ir, Ru, Re,  Rh, Hg, Hf),  and restricted the maximum atomic number of participating species to 83. Further, we also removed strongly ionic compounds formed by halides, oxides, and hydrides, thereby bringing down the total number of considered compounds to 9898. Since our focus is on the phonon thermal transport in semiconductors and insulators, we next removed all compounds with electronic bandgap lower than $0.2$ eV (as obtained from GGA-based DFT in the Materials Project) and limited our study to thermodynamically stable compounds with energy above the convex hull (with respect to all reported materials in the Materials Project) less than $0.02$ eV/atom, thus, further bringing down the number of compounds to 1451. Finally, considering the $N_{atom}^4$ scaling of computational cost for thermal conductivity calculations \cite{mcgaughey2019}, we restricted to small unitcell compounds with up to 15 atoms in the unitcell. Of the total of 429 filtered compounds, during structure relaxation and dynamical stability calculations, 204 compounds are further dropped due to the presence of imaginary phonon modes and/or non-convergence of self-consistent functional calculations, and the full thermal conductivity calculations are carried out for 225 compounds. 

The diversity, stability, and electronic bandgaps of shortlisted 232 materials on which full thermal conductivity calculations are carried out are presented in Fig.~\ref{fig_screening}(b). The selected materials vary across 32 spacegroups and have 43 participating atomic species. Around $2/3^{rd}$ of the compounds are from cubic and tetragonal crystal families, while the remaining compounds are almost equally distributed amongst other crystal families. Around half of the compounds are chalcogenides and around 10\% are intermetallics. 
More than 70\% of considered compounds have energy above the convex hull of 0 eV/atom and are stable for decomposition into any other compounds available on the Materials Project. All considered compounds have negative formation energies, and 70\% compounds have electronic bandgaps lower than $1.5$ eV.

\begin{figure}
\begin{center}
\epsfbox{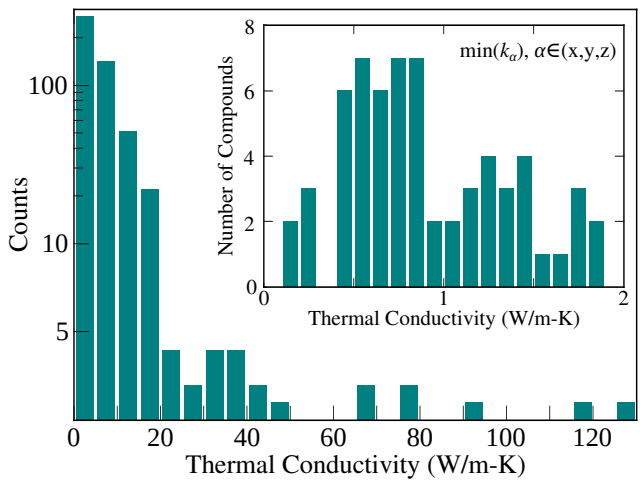}
\end{center}
\caption{The thermal conductivity distribution of considered materials. For non-cubic materials, the non-equivalent directions are considered as different data points (i.e., number of datapoints are more than the number of compounds in the main figure). The inset reports distribution of minimum directional thermal conductivity for low thermal conductivity materials. Forty materials are identified to have thermal conductivities lower than $1.0$ $\text{W/m-K}$ in at least one orthogonal direction at 300 K.}
\label{fig_thermal_K_description}
\end{figure}

The thermal conductivities of considered compounds at a temperature of 300 K are reported in Fig.~\ref{fig_thermal_K_description}. While the thermal conductivity varies three orders of magnitude between $0.16$ and $114$ $\text{W/m-K}$ for considered compounds, around half of the compounds have predicted average thermal conductivity, $k_{avg} = \frac{k_x + k_y + k_z}{3}$, lower than 5 $\text{W/m-K}$,  80\% of the compounds have $k_{avg} < 10$  $\text{W/m-K}$, and only nine considered compounds have $k_{avg} > 25$ $\text{W/m-K}$. Other than $\text{LiCB}$, which has all light atoms and hence high $k_{avg}$,  $k_{avg} > 25$ $\text{W/m-K}$ is obtained only for pnictogen-based compounds, whereas low thermal conductivity of $k_{avg} < 0.5$ $\text{W/m-K}$ is obtained only for chalcogenides and intermetallic compounds (see Table~\ref{table_k}). For low thermal conductivity applications, such as thermoelectric energy generation, thermal insulation, etc., 12 materials are obtained with a thermal conductivity lower than  $0.5$ $\text{W/m-K}$ in at least one of the three orthogonal directions (listed in Table~\ref{table_k}).

\begin{table}
\caption{Selected materials for which (a) directional-average thermal conductivity is lower than $0.5$ $\text{W/m-K}$, (b) minimum directional thermal conductivity is lower than $0.5$ $\text{W/m-K}$, and (c) directional average thermal conductivity is larger than $25$ $\text{W/m-K}$ at 300 K.}
\begin{tabular}{l | c | c}
\hline
Material & Spacegroup Number & $k$ [$\text{W/m-K}$] \\
\hline 
\multicolumn{3}{c}{} \\
\multicolumn{3}{c}{(a) $k_{avg}$ for materials with $k_{avg} < 0.5$ $\text{W/m-K}$} \\
\hline \hline
$\text{SbRbK}_2$ & 225 & $0.16$ \\
$\text{BiCsK}_2$ & 225 & $0.17$ \\
$\text{SbCsK}_2$ & 225 & $0.23$ \\
$\text{AgKSe}$ & 129 & $0.33$ \\
$\text{SbRbNa}_2$ & 225 & $0.43$ \\
$\text{SbTlLi}_2$ & 225 & $0.47$ \\
\hline 
\multicolumn{3}{c}{} \\
\multicolumn{3}{c}{(b) $k_{min}$ for materials with $k_{min} < 0.5$ $\text{W/m-K}$} \\
\hline \hline
$\text{SbRbK}_2$ & 225 & $0.16$ \\
$\text{BiCsK}_2$ & 225 & $0.17$ \\
$\text{SbCsK}_2$ & 225 & $0.23$ \\
$\text{CsAs}_3\text{Cd}_4$ & 123 & $0.25$ \\
$\text{AgKSe}$ & 129 & $0.28$ \\
$\text{RbAs}_3\text{Zn}_4$ & 123 & $0.41$ \\
$\text{AgScSe}_2$ & 164 & $0.42$ \\
$\text{SbRbNa}_2$ & 225 & $0.43$ \\
$\text{PbSnS}_2$ & 26 & $0.45$ \\
$\text{SbTlLi}_2$ & 225 & $0.47$ \\
$\text{AgAsS}$ & 62 & $0.49$ \\
$\text{BiNaSe}_2$ & 166 & $0.50$ \\
\hline
\multicolumn{3}{c}{} \\
\multicolumn{3}{c}{(c) $k_{avg}$ for materials with $k_{avg} > 25$ $\text{W/m-K}$} \\
\hline \hline
$\text{AlGaN}_2$ & 115 & $114$ \\
$\text{LiCB}_2$ & 194 & $87.1$ \\
$\text{GaAl}_3\text{N}_4$ & 215 & $77.7$ \\
$\text{AlGa}_3\text{N}_4$ & 215 & $66.6$ \\
$\text{AlGaP}_2$ & 115 & $42.8$ \\
$\text{SnZnP}_2$ & 115 & $39.6$ \\
$\text{InPAs}_2$ & 115 & $38.9$ \\
$\text{NiSi}_3\text{P}_4$ & 121 & $35.0$ \\
$\text{CaAs}_2\text{Be}_2$ & 164 & $27.5$ \\
\hline
\end{tabular}
\label{table_k}
\end{table}

\begin{figure}
\begin{center}
\epsfbox{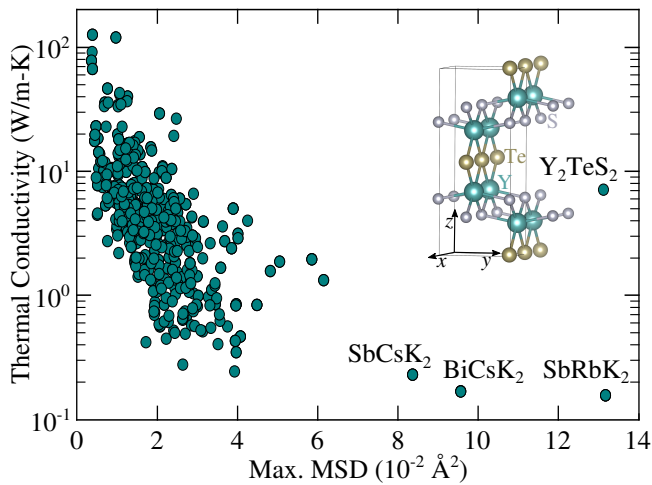}
\end{center}
\caption{The directional thermal conductivity of materials reported against the directional thermal mean square displacement of atoms. The considered thermal mean square displacements (MSD) are maximum over all atoms in the material. The inset shows the atomic structure of anomalous $\text{Y}_2\text{TeS}_2$ for which thermal conductivity is high despite the large thermal MSD of atoms.}
\label{fig_RMSD}
\end{figure}

According to Eqn.~\ref{eqn_theory_conduct}, the phonon thermal conductivity of a material is dependent on phonon group velocity, heat capacity, and scattering lifetimes. The phonon group velocity and heat capacity are large in strongly bonded materials with high bond stiffness. Further, atoms move little around their equilibrium position for strongly bonded materials and predominantly sample the harmonic part of the potential energy surface.  As such, the phonon scattering lifetimes are also large (low phonon-phonon scattering) in strongly bonded  materials. This strong vs. weak bonding of materials is characterizable using the thermal mean square displacements (MSD) of atoms around their equilibrium positions.
The thermal MSD of atoms for considered compounds are plotted in Fig.~\ref{fig_RMSD}. In particular, we plot the directional thermal conductivity against the maximum directional MSD of atoms in Fig.~\ref{fig_RMSD}. 

As can be seen from Fig.~\ref{fig_RMSD}, the thermal conductivity of considered compounds correlates strongly (negative correlation) with thermal MSD, and, in general, materials with small (large) MSD result in a high (low) thermal conductivity. The MSD of Cs atoms in $\text{SbCsK}_2$ and $\text{BiCsK}_2$, of Rb atoms in $\text{SbRbK}_2$ and that of Te atoms in $\text{Y}_2\text{TeS}_2$ are larger than $0.08$ $\text{\AA}^2$ at 300 K. Such large MSD are also observed in cage-like compounds, such as clathrates \cite{sales2001}, and indicate rattler-like motion of concerned atoms resulting in a  scattering of heat-carrying acoustic phonons and hence low material thermal conductivity \cite{sales2001,bentien2004,jain2020,godse2022}. This is indeed the case for $\text{SbCsK}_2$, $\text{BiCsK}_2$, and $\text{SbRbK}_2$, and the predicted thermal conductivity of all three of these compounds is lower than 0.3  $\text{W/m-K}$. For $\text{Y}_2\text{TeS}_2$, however, the thermal conductivity is high at 7.1 $\text{W/m-K}$ despite the large MSD of $0.13$ $\text{\AA}^2$. This anomaly in thermal transport in $\text{Y}_2\text{TeS}_2$ is related to its unique crystal structure (reported in the inset of Fig.~\ref{fig_RMSD}) where the Y and S atoms form a layered structure and the individual layers of $\text{YS}$ are coupled to each other through Te atoms. The bonding of Te atoms with neighboring YS layers is planar ($xz$-plane in the structure shown in the inset Fig.~\ref{fig_RMSD}). As such, Te atoms are non-bonded in the $y$-direction, resulting in their large MSD. However, even with this non-bonding of Te atoms in the $y$-direction, the heat is still able to flow through covalent-bonded YS layers, thereby resulting in a relatively large thermal conductivity even with a large MSD of atoms.

\begin{figure}
\begin{center}
\epsfbox{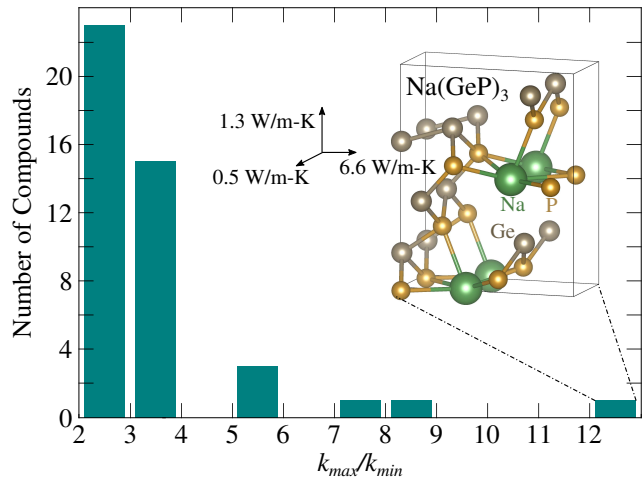}
\end{center}
\caption{The thermal transport anisotropy distribution of highly anisotropic covalently bonded materials. The highest obtained anisotropy is $12.6$ for $\text{Na(GeP)}_3$ and 20 materials are identified to have thermal transport anisotropy larger than the highest literature reported value of $3.1$ \cite{jain2015}.}
\label{fig_aniso}
\end{figure}

Moving further, we next compute thermal transport anisotropy as $k_{max}/k_{min}$ where $k_{max}$, $k_{min}$ are maximum and minimum directional thermal conductivities and report the results in Fig.~\ref{fig_aniso}. We considered only covalently bonded compounds and report only compounds with anisotropy larger than two in Fig.~\ref{fig_aniso}. In total, we have identified 44  compounds with thermal anisotropy larger than two and six compounds with anisotropy larger than five. The highest anisotropy obtained is of $12.6$ for $\text{Na(GeP)}_3$ in spacegroup 26 with thermal conductivity values of $0.52$, $6.60$, and $1.26$ $\text{W/m-K}$ along the three orthogonal directions. Amongst covalently bonded materials, to the best of our knowledge, the highest experimentally measured thermal transport anisotropy is reported by Liu et al.~\cite{liu2020} with a value of two for titanium trisulphide in the monoclinic ($\text{P}2_1/\text{m}$) spacegroup. Computationally, black phosphorene is reported to have high anisotropy of $3.1$ in the basal plane at room temperature \cite{jain2015}. Interestingly we have identified 20 covalently bonded materials here with thermal transport anisotropy larger than $3.1$.

\section*{Discussion}
\label{sec_conc}
With recent advances in computational resources, DFT-driven material property databases are now emerging with millions of entries for relatively simple material properties, such as thermodynamic stability, electronic bandgap, and mobility \cite{materialsProject, oqmd}. The availability of such databases is facilitating the data-driven accelerated discovery of novel materials, for instance, using machine learning techniques. Owing to the high computational cost of phonon thermal conductivity prediction (10000-100000 cpu-hours compared to only 1-10 cpu-hours for simpler properties), however,  the available data on phonon thermal conductivity of materials is scarce; limited majorly to monoatomic and diatomic compounds with simple crystal structures \cite{xia2020b, miyazaki2021, he2021, pal2021}. Further, the available data is obtained using different computational packages employing diverse methodologies/simulation parameters \cite{lindsay2018,mcgaughey2019}. In the absence of high-quality consistent data, the data-driven approaches are not of much use and it is not possible to extract useful insights from available data using data-driven approaches. 

The data presented here is highly curated and is obtained using similar simulation settings for all compounds. Further, the data is sufficiently diverse to cover different phonon thermal transport physics present in varying material systems. The computed data could be readily used for data-driven materials discovery using machine learning-based methods. In this regard, we have made available all thermal conductivity data, including the atomic structures, simulation parameters, atomic MSD, and final thermal conductivity in the raw format. Further, the intermediate computational data, such as extracted harmonic and anharmonic force constants, and mode-level phonon properties of all compounds, could also be made available upon a reasonable request. Besides identifying 40 low thermal conductivity and 20 high anisotropy compounds, we hope that the generated data will pave the way forward for data-driven exploration of material thermal properties.

\section*{Conclusions}
In summary, we performed ab-initio driven high-throughput DFT computations to find materials with low phonon thermal conductivity and high thermal transport anisotropy. We solved the Boltzmann transport equation for 225 stable ternary semiconductors and identified 40 compounds with a thermal conductivity lower than 1 $\text{W/m-K}$ and 21 materials with thermal transport anisotropy greater than $3.0$. Since all of our calculations are performed systematically using the same computational tools and similar computational parameters, the presented data can be readily used for the data-driven discovery of new thermal transport physics and/or new materials with desired thermal properties.

\section*{Methods}
\label{sec_theory}

We calculate the thermal conductivity, $k_{\alpha}$, of semiconducting solid in the $\alpha$-direction by solving the Boltzmann transport equation and using the Fourier's law as \cite{reissland1973}:
\begin{equation}
 \label{eqn_theory_conduct}
 k_{\alpha} = \sum_{{\boldsymbol{q}}} \sum_{\nu} c_{{\boldsymbol{q}}\nu} v_{{\boldsymbol{q}}\nu, \alpha}^{2} \tau_{{\boldsymbol{q}}\nu, \alpha}.
\end{equation}
The summation in the Eqn.~\ref{eqn_theory_conduct} is over all the phonon wavevectors, ${\boldsymbol{q}}$, and polarizations, $\nu$ and  $c_{{\boldsymbol{q}}\nu}$ is the phonon specific heat, $v_{{\boldsymbol{q}}\nu,\alpha}$ is the $\alpha$ component of phonon group velocity vector ${\boldsymbol{v}_{{\boldsymbol{q}}\nu}}$, and $\tau_{{\boldsymbol{q}}\nu, \alpha}$ is the phonon scattering time. Phonons are bosons and follow the Bose-Einstein distribution, when in equilibrium. The phonon specific heat can be obtained from the phonon vibrational frequencies as:
\begin{equation}
\label{eqn_theory_cph}
c_{{\boldsymbol{q}}\nu} = \frac{\hbar\omega_{{\boldsymbol{q}}\nu}}{V} \frac{\partial n^{o}_{{\boldsymbol{q}}\nu}}{\partial T} = \frac{k_\text{B} x^2 e^x }{(e^x-1)^2}.
\end{equation}
The $n^o_{{\boldsymbol{q}}\nu}$ in Eqn.~\ref{eqn_theory_cph} is the Bose-Einstein distribution ($n^o_{{\boldsymbol{q}}\nu} = \frac{1}{e^x-1}$), $\hbar$ is the reduced Planck constant, $\omega_{{\boldsymbol{q}}\nu}$ is the phonon frequency, $V$ is the crystal volume,  $T$ is the temperature, $k_\text{B}$ is the Boltzmann constant, and $x = \frac{\hbar\omega_{{\boldsymbol{q}}\nu}}{k_\text{B}T}$. The phonon group velocities are obtained as:
\begin{equation}
    {\boldsymbol{v}_{{\boldsymbol{q}}\nu}} = \frac{\partial \omega_{\boldsymbol{q}\nu}}{\partial \boldsymbol{q}}.
\end{equation}
The phonon frequencies are obtained from the diagonalization of the dynamical matrix as:
\begin{equation}
 \label{eqn_theory_evproblem}
 \omega_{{\boldsymbol{q}}\nu}^{2} {\boldsymbol{e}}_{{\boldsymbol{q}}\nu} = {\boldsymbol{D}}_{{\boldsymbol{q}}} \cdot {\boldsymbol{e}}_{{\boldsymbol{q}}\nu},
\end{equation}
where ${\boldsymbol{e}}_{{\boldsymbol{q}}\nu}$ is the eigenvector and ${\boldsymbol{D}}_{{\boldsymbol{q}}}$ is the Dynamical matrix  whose elements, $D_{\boldsymbol{q}}^{3(b-1)+\alpha, 3(b^{'}-1) + \beta}$, given by:
\begin{equation}
 \label{eqn_theory_dynamical}
 \begin{split}
 D_{\boldsymbol{q}}^{3(b-1)+\alpha, 3(b^{'}-1) + \beta} = \frac{1}{\sqrt{m_b m_{b^{'}}}} \sum_{l^{'}} \Phi_{b0;b^{'}l^{'}}^{\alpha\beta} 
 \\
 \exp{\{i[{\boldsymbol{q}}.( {\boldsymbol{r}}_{b^{'}l^{'}}  -  {\boldsymbol{r}}_{b0}  )] \}},
 \end{split}
\end{equation}
where the summation is over all unit-cells in the lattice. Here $m_b$ is the mass of atom $b$ in the unit-cell,  $\boldsymbol{r}_{bl}$ is the position vector of atom $b$ in the $l^{th}$ unit-cell, and $\Phi_{ij}^{\alpha\beta}$ is the real-space ($ij, \alpha\beta$)-element of the  harmonic force constant matrix $\boldsymbol{\Phi}$.
The phonon scattering time is obtained by considering the three-phonon scattering processes as\cite{reissland1973, wallace1972, jain2020}:

\begin{equation}
\begin{split}
 \label{eqn_rta_3ph}
 \frac{1}{\tau_{\boldsymbol{q}\nu}^{}} 
 =
  \sum_{{\boldsymbol{q}_{1}}\nu_{1}}
 \sum_{{\boldsymbol{q}_{2}}\nu_{2}}
 \bigg\{
 \Big\{
  {(n_{{\boldsymbol{q}_{1}\nu_{1}}}  - n_{{\boldsymbol{q}_{2}\nu_{2}}})}
 W^{+}
  \Big\} 
  +
  \\
  \frac{1}{2}
   \Big\{
  (n_{{\boldsymbol{q}_{1}\nu_{1}}} + n_{{\boldsymbol{q}_{2}\nu_{2}}} +  1)
  W^{-}
   \Big\}
   \bigg\},
 \end{split}
 \end{equation}
where $\boldsymbol{W}$ represents scattering probability matrix given by:
\begin{equation}
    \begin{split}
    \label{eqn_W_3ph}
W^{\pm}       
=
\frac{2\pi}{\hbar^2}
 \left|
 \Psi_{ {\boldsymbol{q}} (\pm{\boldsymbol{q}_{1}}) (-{\boldsymbol{q}_{2}}) }^{\nu \nu_{1} \nu_{2}}
 \right|^2
  \delta({\omega_{{{\boldsymbol{q}}_{}}\nu_{}} \pm \omega_{{{\boldsymbol{q}}_{1}}\nu_{1}} - \omega_{{{\boldsymbol{q}}_{2}}\nu_{2}}}).
    \end{split}
\end{equation}
The $\Psi_{ {\boldsymbol{q}} {\boldsymbol{q}_{1}} {\boldsymbol{q}_{2}} }^{\nu \nu_{1} \nu_{2}}$ are the Fourier transform of real-space cubic constants, $\Psi^{\alpha\beta\gamma}_{bl;b^{'} l^{'};b^{''} l^{''}}$, and are obtained as:
\begin{equation}
    \begin{split}
        \label{eqn_cubic_IFC}
\Psi_{ {\boldsymbol{q}} {\boldsymbol{q}_{1}} {\boldsymbol{q}_{2}} }^{\nu \nu_{1} \nu_{2}}
=
 \Psi_{ {\boldsymbol{q}} {\boldsymbol{q}^{'}} {\boldsymbol{q}^{''}} }^{\nu \nu^{'} \nu^{''}} = 
 N
 {\left(\frac{\hbar}{2N}\right)}^{\frac{3}{2}}
 \sum_{b} \sum_{b^{'} l^{'}}
\sum_{b^{''} l^{''}} 
\sum_{\alpha\beta\gamma} 
\Psi^{\alpha\beta\gamma}_{bl;b^{'} l^{'};b^{''} l^{''}}
\\
\times
\frac{
{{\tilde{e}}_{b,\boldsymbol{q}\nu}^{\alpha}}  
{{\tilde{e}}_{b^{'},{\boldsymbol{q}}^{'} \nu^{'}}^{\beta}} 
{{\tilde{e}}_{b^{''},{\boldsymbol{q}}^{''} \nu^{''}}^{\gamma}} }{\sqrt{ 
{m_b \omega_{\boldsymbol{q}\nu}}  
{m_{b^{'}} \omega_{{\boldsymbol{q}}^{'}\nu^{'}}}   
{m_{b^{''}} \omega_{{\boldsymbol{q}}^{''}\nu^{''}}}   }}  
e^{[i( {{\boldsymbol{q}}^{'}}  \cdot{\boldsymbol{r}}_{0l^{'}} 
+    {{\boldsymbol{q}}^{''}}  \cdot{\boldsymbol{r}}_{0l^{''}} )]} ,
    \end{split}
\end{equation}
The $\delta$ in Eqn.~\ref{eqn_W_3ph} represents the delta-function ensuring energy conservation and the summation in Eqn.~\ref{eqn_cubic_IFC} is performed over phonon wavevectors satisfying crystal momentum conservation, i.e., $\boldsymbol{q} + \boldsymbol{q_1} + \boldsymbol{q_2} = \boldsymbol{G}$, where $\boldsymbol{G}$ is the reciprocal space lattice vector.

The computation of phonon thermal conductivity via Eqns.~\ref{eqn_theory_conduct} requires harmonic and anharmonic interatomic force constants. While the anharmonic force constants can be obtained from the finite-difference of density functional theory (DFT) forces by displacing one or more atoms in the computational supercell corresponding to the equilibrium positions of atoms, the force constants obtained from this approach samples the PES at 0 K. For moderately anharmonic solids, this 0 K sampling is adequate and results in a minimal error. For strongly anharmonic solids, the PES anharmonicity is strongly dependent on atomic displacements and is a strong function of temperature. In this work, T-dependent force constants are obtained by force-displacement data fitting on thermally populated supercells. 

The harmonic ($\Phi_{ij}^{\alpha\beta}$), cubic ($\Psi_{ijk}^{\alpha\beta\gamma}$), and quartic ($\Xi_{ijkl}^{\alpha\beta\gamma\delta}$) force constants are obtained by fitting the over-determined force-displacement dataset using the least-square fit. The force acting on atom $i$ in the $\alpha$-direction, $F_i^{\alpha}$, is written in terms of interatomic forceconstants as:
\begin{equation}
\label{eqn_force}
    F_i^{\alpha} = -\frac{\partial U}{\partial u_{i}^{\alpha}},
\end{equation}
where potential energy, $U$ can be expanded using Taylor series as:
\begin{equation}
\begin{split}
\label{eqn_theory_pot}
 U = U_0 + 
 \displaystyle\sum_{i} {\Pi_{i}^{\alpha} u_i^{\alpha}} +
\frac{1}{2!}\displaystyle\sum_{ij} {\Phi_{ij}^{\alpha\beta}u_i^{\alpha}u_j^{\beta}} +
\\
\frac{1}{3!}\displaystyle\sum_{ijk} {\Psi_{ijk}^{\alpha\beta\gamma} u_i^{\alpha}u_j^{\beta}u_k^{\gamma}} +
\\
\frac{1}{4!}\displaystyle\sum_{ijkl} {\Xi_{ijkl}^{\alpha\beta\gamma\delta} u_i^{\alpha}u_j^{\beta}u_k^{\gamma}u_l^{\delta}} +
O\left({u^5}\right),
\end{split}
\end{equation}
where $u_i^{\alpha}$ is the displacement of atom $i$ in the $\alpha$-direction from its equilibrium position. The force-displacement dataset is generated using a stochastic technique in which structures are thermal populated corresponding to the given temperature by displacing atoms according to the distribution \cite{west2006, hellman2013, shulumba2017}:
\begin{equation}
    \begin{split}
        \label{eqn_thermal_population}
        u_{b,l}^{\alpha} = \frac{1}{\sqrt{N}}
        \sum_{\boldsymbol{q}\nu}
        \sqrt{\frac{\hbar(n_{\boldsymbol{q}\nu} + 1)}
            {m_b\omega_{\boldsymbol{q}\nu}}}
            \cos{(2\pi\eta_{1, \boldsymbol{q}\nu})}
            \\
            \sqrt{-\ln{(1-\eta_{2, \boldsymbol{q}\nu})}}
            {{\tilde{e}^\alpha}_{b, \boldsymbol{q}\nu}}
            e^{i \boldsymbol{q}  \cdot{\boldsymbol{r}}_{0l} },
    \end{split}
\end{equation}
where $\eta_{1, \boldsymbol{q}\nu}$ and $\eta_{2, \boldsymbol{q}\nu}$ are random numbers sampled from a uniform distribution and constrained by $\eta_{1, \boldsymbol{q}\nu} = \eta_{1, -\boldsymbol{q}\nu}$ and $\eta_{2, \boldsymbol{q}\nu} = \eta_{2, -\boldsymbol{q}\nu}$, and forces acting on atoms in these thermally populated stuctures are obtained using the DFT calculations. The phonon frequencies and eigenvectors needed in Eqn.~\ref{eqn_thermal_population} are obtained from harmonic force constants calculated using density functional perturbation theory (DFPT) calculations including the effect of LO-TO splitting.  The temperature is taken into account in Eqn.~\ref{eqn_thermal_population} via T-dependent mode population ($n_{\boldsymbol{q}\nu}$). During the data-fitting step for extraction of anharmonic force constants using Eqns.~\ref{eqn_force} and \ref{eqn_theory_pot}, the harmonic force constants contribution (including the contribution from non-zero Born effective charges) is removed from the forces and only cubic and quartic force constants are fitted (simultaneously) to the residual forces-displacement data.

The DFT calculations are performed using the openly available quantum mechanical simulation package Quantum Espresso employing planewave-based basis set with norm-conserving Vanderbilt pseudopotentials \cite{giannozzi2009}. The planewave kinetic energy cutoff is set at 80 Ry in all calculations. The structure relaxations are performed using primitive unitcells and the electronic Brillouin zone is sampled using Monkhorst-Pack wavevector grid of size $k_i$ such that $k_i.|a_i^{pri}| \sim 30$ $\text{\AA}$, where $|a_i^{pri}|$ represents the length of primitive unitcell lattice vector $\textbf{a}_i^{pri}$. The electronic total energy is converged to within $10^{-10}$ Ry/atom during self-consistent cycles and the structure relaxations are performed with force convergence criterion of $10^{-5}$ Ry/$\text{\AA}$.

The harmonic force constants and Born effective charges are obtained using DFPT calculations on primitive unitcells. The DFPT calculations are initially performed on phonon wavevector grids of size $q_i^c$ such that $q_i^c.|a_i^{pri}| \sim 30$ $\text{\AA}$ and are later interpolated to grids of size $q_i^f$ such that $q_i^f.|a_i^{pri}| \sim 100$ $\text{\AA}$ for three-phonon scattering calculations. The cubic and quartic IFCs are obtained from Taylor-series fitting of Hellmann-Feynman forces obtained on 200 thermally populated  supercells  (corresponding to a temperature of 300 K) obtained from $N_i$ repetitions of the conventional unitcell such that $N_i.|a_i^{conv}| \sim 15$.  The cubic and quartic force constant interaction cutoffs are set at $6.5$ and $4.0$ $\text{\AA}$ for majority of compounds though for some compounds with lower symmetries, these values are reduced to $5.0$ and $2.5$ $\text{\AA}$. The actual values of simulation parameters are provided in the Table~\ref{table_simulation_parameters} for all compounds.

\section*{Acknowledgement}
The author acknowledges the financial support from IRCC-IIT Bombay and the National Supercomputing Mission, Government of India (Grant Number: DST/NSM/R\&D-HPC-Applications/2021/10).  The calculations are carried out on SpaceTime-II supercomputing facility of IIT Bombay and PARAM Sanganak supercomputing facility of IIT Kanpur.

\newpage
\begin{center}
\begin{longtable*}{l | p{2.6cm} | p{2cm} | p{1.3cm} | p{1.8cm} | p{2cm} | p{1.2cm} | p{0.7cm} | p{0.7cm} | p{3cm} }
\caption{\label{table_simulation_parameters}The simulation parameters and thermal conductivities of compounds for which full DFT calculations are performed and the Boltzmann Transport Equation is solved. The compound names are written as stoichiomtery-spacegroup.} \\ 
\hline
S.No. & Name & $|a^{pri}_i|$ &  $k_i$, $q_i^c$ & $q_i^f$ & $|a^{conv}_i|$ & $N_i$ & \multicolumn{2}{c|}{cutoff} & [$k_x$, $k_y$, $k_z$] \\ 
 &  & &  & & & & $3^{rd}$ & $4^{th}$ & [W/m-K] \\ 
\hline
1  &  $\text{Al}\text{In}\text{Sb}_{2}$-115 & [ 4.5, 4.5, 6.4] & [ 7, 7, 5] & [ 22, 22, 16] & [ 4.5, 4.5, 6.4] & [ 3, 3, 2] & 6.0 & 4.0 & [  18.8,  18.8,  14.8] \\ 
2  &  $\text{Ba}_{2}\text{Cu}_{4}\text{S}_{4}$-139 & [ 6.9, 6.9, 6.9] & [ 4, 4, 4] & [ 14, 14, 14] & [ 3.9, 3.9,12.6] & [ 4, 4, 1] & 6.3 & 4.0 & [   1.3,   1.3,   1.1] \\ 
3  &  $\text{Ca}_{2}\text{Mg}_{4}\text{Sb}_{4}$-164 & [ 4.7, 4.7, 7.5] & [ 6, 6, 4] & [ 22, 22, 14] & [ 8.1, 4.7, 7.5] & [ 2, 3, 2] & 6.5 & 4.0 & [   5.2,   5.1,   4.8] \\ 
4  &  $\text{S}_{4}\text{Te}_{2}\text{Y}_{4}$-71 & [ 7.6, 7.6, 7.6] & [ 4, 4, 4] & [ 14, 14, 14] & [ 4.2, 5.4,13.5] & [ 4, 3, 1] & 6.5 & 4.0 & [   7.6,   7.1,   5.3] \\ 
5  &  $\text{Ba}_{2}\text{Cd}_{4}\text{P}_{4}$-164 & [ 4.5, 4.5, 7.6] & [ 7, 7, 4] & [ 22, 22, 14] & [ 7.7, 4.5, 7.6] & [ 2, 3, 2] & 6.5 & 4.0 & [   5.3,   5.3,   4.4] \\ 
6  &  $\text{As}_{2}\text{Ca}_{2}\text{Rb}_{2}$-129 & [ 5.2, 5.2, 8.0] & [ 6, 6, 4] & [ 20, 20, 12] & [ 5.2, 5.2, 8.0] & [ 3, 3, 2] & 6.5 & 4.0 & [   1.9,   1.9,   0.7] \\ 
7  &  $\text{K}_{2}\text{Li}_{2}\text{S}_{2}$-129 & [ 4.3, 4.3, 7.0] & [ 7, 7, 4] & [ 24, 24, 14] & [ 4.3, 4.3, 7.0] & [ 3, 3, 2] & 6.5 & 4.0 & [   3.0,   3.0,   1.8] \\ 
8  &  $\text{Ni}_{4}\text{Sn}_{4}\text{Zr}_{4}$-216 & [ 4.4, 4.4, 4.4] & [ 7, 7, 7] & [ 24, 24, 24] & [ 6.2, 6.2, 6.2] & [ 2, 2, 2] & 6.0 & 3.0 & [   7.7,   7.7,   7.7] \\ 
9  &  $\text{Ba}_{2}\text{N}_{4}\text{Zr}_{2}$-129 & [ 4.2, 4.2, 8.4] & [ 7, 7, 4] & [ 24, 24, 12] & [ 4.2, 4.2, 8.4] & [ 4, 4, 2] & 6.0 & 3.0 & [   5.6,   5.6,   2.4] \\ 
10  &  $\text{Bi}_{4}\text{Li}_{4}\text{Mg}_{4}$-216 & [ 4.8, 4.8, 4.8] & [ 6, 6, 6] & [ 22, 22, 22] & [ 6.8, 6.8, 6.8] & [ 2, 2, 2] & 6.5 & 4.0 & [   5.4,   5.4,   5.4] \\ 
11  &  $\text{Ag}_{2}\text{Te}_{4}\text{Y}_{2}$-113 & [ 7.1, 7.1, 4.7] & [ 4, 4, 6] & [ 14, 14, 22] & [ 7.1, 7.1, 4.7] & [ 2, 2, 3] & 6.5 & 4.0 & [   0.7,   0.7,   0.6] \\ 
12  &  $\text{Cu}_{3}\text{Se}_{4}\text{Ta}$-215 & [ 5.7, 5.7, 5.7] & [ 5, 5, 5] & [ 18, 18, 18] & [ 5.7, 5.7, 5.7] & [ 3, 3, 3] & 6.5 & 4.0 & [   4.3,   4.3,   4.2] \\ 
13  &  $\text{Ag}_{4}\text{As}_{4}\text{S}_{4}$-62 & [ 5.9, 4.1,12.0] & [ 5, 7, 3] & [ 18, 24,  8] & [ 5.9, 4.1,12.0] & [ 3, 4, 1] & 6.0 & 4.0 & [   0.6,   1.0,   0.5] \\ 
14  &  $\text{Ge}_{2}\text{Se}_{8}\text{Zn}_{4}$-121 & [ 6.7, 6.7, 6.7] & [ 4, 4, 4] & [ 16, 16, 16] & [ 5.8, 5.8,10.7] & [ 3, 3, 1] & 5.3 & 4.0 & [   5.4,   5.4,   8.8] \\ 
15  &  $\text{Cd}_{2}\text{Ga}_{4}\text{Te}_{8}$-82 & [ 7.5, 7.5, 7.5] & [ 4, 4, 4] & [ 14, 14, 14] & [ 6.3, 6.3,12.0] & [ 2, 2, 1] & 6.0 & 4.0 & [   3.3,   3.3,   4.0] \\ 
16  &  $\text{As}_{8}\text{In}_{8}\text{Sr}_{4}$-194 & [ 4.3, 4.3,18.3] & [ 7, 7, 2] & [ 24, 24,  6] & [ 4.3, 7.5,18.3] & [ 3, 2, 1] & 6.5 & 4.0 & [   4.9,   4.9,   2.8] \\ 
17  &  $\text{Li}_{4}\text{Sb}_{4}\text{Zn}_{4}$-216 & [ 4.5, 4.5, 4.5] & [ 7, 7, 7] & [ 22, 22, 22] & [ 6.4, 6.4, 6.4] & [ 2, 2, 2] & 6.0 & 3.0 & [   6.0,   6.0,   6.0] \\ 
18  &  $\text{K}_{2}\text{Li}_{2}\text{Se}_{2}$-129 & [ 4.5, 4.5, 7.3] & [ 7, 7, 4] & [ 22, 22, 14] & [ 4.5, 4.5, 7.3] & [ 3, 3, 2] & 6.5 & 4.0 & [   4.7,   4.7,   2.4] \\ 
19  &  $\text{Na}_{6}\text{Sb}_{6}\text{Se}_{12}$-166 & [ 4.1, 4.1,21.2] & [ 7, 7, 1] & [ 24, 24,  6] & [ 7.1, 4.1,21.2] & [ 2, 4, 1] & 6.5 & 4.0 & [   1.3,   1.3,   0.6] \\ 
20  &  $\text{Bi}_{4}\text{Ca}_{4}\text{Li}_{4}$-62 & [ 7.8, 4.7, 8.5] & [ 4, 6, 4] & [ 14, 22, 12] & [ 7.8, 4.7, 8.5] & [ 2, 3, 2] & 6.0 & 3.0 & [   3.5,   3.5,   2.7] \\ 
21  &  $\text{Cd}\text{Se}_{2}\text{Zn}$-115 & [ 4.2, 4.2, 6.0] & [ 7, 7, 5] & [ 24, 24, 18] & [ 4.2, 4.2, 6.0] & [ 4, 4, 2] & 6.0 & 4.0 & [   7.2,   7.2,   6.3] \\ 
22  &  $\text{K}_{8}\text{Rb}_{4}\text{Sb}_{4}$-225 & [ 6.1, 6.1, 6.1] & [ 5, 5, 5] & [ 16, 16, 16] & [ 8.7, 8.7, 8.7] & [ 2, 2, 2] & 6.5 & 4.0 & [   0.2,   0.2,   0.2] \\ 
23  &  $\text{Cd}_{3}\text{Se}_{4}\text{Zn}$-215 & [ 6.1, 6.1, 6.1] & [ 5, 5, 5] & [ 16, 16, 16] & [ 6.1, 6.1, 6.1] & [ 2, 2, 2] & 6.0 & 4.0 & [   5.9,   5.9,   5.9] \\ 
24  &  $\text{Ga}_{4}\text{S}_{8}\text{Zn}_{2}$-121 & [ 6.6, 6.6, 6.6] & [ 5, 5, 5] & [ 16, 16, 16] & [ 5.2, 5.2,10.9] & [ 3, 3, 1] & 5.4 & 4.0 & [  13.9,  13.9,  13.2] \\ 
25  &  $\text{Cd}_{4}\text{Se}\text{Te}_{3}$-25 & [ 4.6, 9.2, 6.5] & [ 7, 3, 5] & [ 22, 12, 16] & [ 4.6, 9.2, 6.5] & [ 3, 2, 2] & 6.5 & 4.0 & [   4.2,   4.0,   3.9] \\ 
26  &  $\text{N}_{4}\text{Sr}_{2}\text{Zr}_{2}$-129 & [ 4.1, 4.1, 8.3] & [ 7, 7, 4] & [ 26, 26, 12] & [ 4.1, 4.1, 8.3] & [ 4, 4, 2] & 6.5 & 4.0 & [   4.9,   4.9,   2.9] \\ 
27  &  $\text{Cd}_{8}\text{Sc}_{16}\text{Se}_{32}$-227 & [ 8.0, 8.0, 8.0] & [ 4, 4, 4] & [ 12, 12, 12] & [11.4,11.4,11.4] & [ 1, 1, 1] & 5.5 & 3.0 & [   4.7,   4.7,   4.7] \\ 
28  &  $\text{Cd}_{4}\text{Se}_{3}\text{Te}$-25 & [ 4.5, 8.9, 6.3] & [ 7, 3, 5] & [ 22, 12, 16] & [ 4.5, 8.9, 6.3] & [ 3, 2, 2] & 6.3 & 4.0 & [   4.7,   4.3,   4.2] \\ 
29  &  $\text{Bi}_{4}\text{Cs}_{4}\text{K}_{8}$-225 & [ 6.3, 6.3, 6.3] & [ 5, 5, 5] & [ 16, 16, 16] & [ 9.0, 9.0, 9.0] & [ 2, 2, 2] & 6.5 & 4.0 & [   0.2,   0.2,   0.2] \\ 
30  &  $\text{Li}_{8}\text{Sb}_{4}\text{Tl}_{4}$-225 & [ 5.0, 5.0, 5.0] & [ 6, 6, 6] & [ 20, 20, 20] & [ 7.0, 7.0, 7.0] & [ 2, 2, 2] & 6.5 & 4.0 & [   0.5,   0.5,   0.5] \\ 
31  &  $\text{Na}_{4}\text{S}\text{Se}$-123 & [ 4.7, 4.7, 6.7] & [ 6, 6, 4] & [ 22, 22, 16] & [ 4.7, 4.7, 6.7] & [ 3, 3, 2] & 6.5 & 4.0 & [   3.4,   3.4,   3.1] \\ 
32  &  $\text{K}_{4}\text{Na}_{4}\text{S}_{4}$-62 & [ 7.8, 4.6, 8.5] & [ 4, 6, 4] & [ 14, 22, 12] & [ 7.8, 4.6, 8.5] & [ 2, 3, 2] & 5.5 & 3.0 & [   1.3,   1.7,   1.3] \\ 
33  &  $\text{Ca}_{4}\text{Ge}_{4}\text{Sr}_{4}$-62 & [ 8.2, 4.9, 9.3] & [ 4, 6, 3] & [ 12, 20, 12] & [ 8.2, 4.9, 9.3] & [ 2, 3, 2] & 5.8 & 3.1 & [   1.4,   1.7,   1.5] \\ 
34  &  $\text{B}_{2}\text{C}_{2}\text{Li}_{2}$-194 & [ 2.7, 2.7, 7.0] & [11,11, 4] & [ 36, 36, 14] & [ 2.7, 2.7, 7.0] & [ 5, 5, 2] & 5.0 & 2.0 & [ 119.3, 119.3,  22.8] \\ 
35  &  $\text{Pb}_{12}\text{Se}_{6}\text{Te}_{6}$-166 & [ 4.5, 4.5,22.1] & [ 7, 7, 1] & [ 22, 22,  6] & [ 7.8, 4.5,22.1] & [ 2, 3, 1] & 6.5 & 4.0 & [   0.6,   0.6,   0.6] \\ 
36  &  $\text{Li}_{4}\text{S}_{8}\text{Ta}_{4}$-194 & [ 3.4, 3.4,13.0] & [ 9, 9, 2] & [ 30, 30,  8] & [ 3.4, 5.8,13.0] & [ 4, 3, 1] & 6.4 & 4.0 & [  15.6,  15.6,   2.0] \\ 
37  &  $\text{Cd}_{3}\text{S}_{4}\text{Zn}$-215 & [ 5.8, 5.8, 5.8] & [ 5, 5, 5] & [ 18, 18, 18] & [ 5.8, 5.8, 5.8] & [ 3, 3, 3] & 6.5 & 4.0 & [  10.8,  10.8,  10.8] \\ 
38  &  $\text{Mg}_{4}\text{Sb}_{4}\text{Sr}_{2}$-164 & [ 4.7, 4.7, 7.8] & [ 6, 6, 4] & [ 22, 22, 14] & [ 8.2, 4.7, 7.8] & [ 2, 3, 2] & 6.5 & 4.0 & [   7.8,   7.7,   7.2] \\ 
39  &  $\text{Ag}_{2}\text{Sc}_{2}\text{Se}_{4}$-164 & [ 3.9, 3.9, 6.6] & [ 8, 8, 5] & [ 26, 26, 16] & [ 6.8, 3.9, 6.6] & [ 2, 4, 2] & 6.5 & 4.0 & [   0.9,   0.8,   0.4] \\ 
40  &  $\text{Cu}_{3}\text{Nb}\text{Te}_{4}$-215 & [ 6.0, 6.0, 6.0] & [ 5, 5, 5] & [ 18, 18, 18] & [ 6.0, 6.0, 6.0] & [ 2, 2, 2] & 6.0 & 3.0 & [   5.7,   5.7,   5.7] \\ 
41  &  $\text{Cu}_{2}\text{Sc}_{2}\text{Se}_{4}$-156 & [ 3.9, 3.9, 6.3] & [ 8, 8, 5] & [ 26, 26, 16] & [ 6.8, 3.9, 6.3] & [ 2, 4, 2] & 6.3 & 4.0 & [   2.9,   3.0,   2.0] \\ 
42  &  $\text{K}_{4}\text{Na}_{8}\text{Sb}_{4}$-225 & [ 5.5, 5.5, 5.5] & [ 5, 5, 5] & [ 18, 18, 18] & [ 7.8, 7.8, 7.8] & [ 2, 2, 2] & 6.5 & 4.0 & [   0.8,   0.8,   0.8] \\ 
43  &  $\text{Bi}\text{Ca}_{3}\text{N}$-221 & [ 4.9, 4.9, 4.9] & [ 6, 6, 6] & [ 20, 20, 20] & [ 4.9, 4.9, 4.9] & [ 3, 3, 3] & 6.5 & 4.0 & [   4.8,   4.8,   4.8] \\ 
44  &  $\text{Li}_{2}\text{Rb}_{2}\text{Se}_{2}$-129 & [ 4.6, 4.6, 7.6] & [ 6, 6, 4] & [ 22, 22, 14] & [ 4.6, 4.6, 7.6] & [ 3, 3, 2] & 6.5 & 4.0 & [   2.0,   2.0,   1.0] \\ 
45  &  $\text{Ni}_{4}\text{Sb}_{4}\text{Y}_{4}$-216 & [ 4.5, 4.5, 4.5] & [ 7, 7, 7] & [ 22, 22, 22] & [ 6.3, 6.3, 6.3] & [ 2, 2, 2] & 6.0 & 4.0 & [   8.0,   8.0,   8.0] \\ 
46  &  $\text{Cu}_{3}\text{Nb}\text{Se}_{4}$-215 & [ 5.7, 5.7, 5.7] & [ 5, 5, 5] & [ 18, 18, 18] & [ 5.7, 5.7, 5.7] & [ 3, 3, 3] & 6.5 & 4.0 & [   4.4,   4.4,   4.4] \\ 
47  &  $\text{Cs}\text{P}_{3}\text{Zn}_{4}$-123 & [ 4.1, 4.1,10.4] & [ 7, 7, 3] & [ 24, 24, 10] & [ 4.1, 4.1,10.4] & [ 4, 4, 1] & 5.2 & 4.0 & [   2.4,   2.4,   0.7] \\ 
48  &  $\text{Cd}_{2}\text{Ga}_{4}\text{Se}_{8}$-82 & [ 6.9, 6.9, 6.9] & [ 4, 4, 4] & [ 16, 16, 16] & [ 5.9, 5.9,11.0] & [ 3, 3, 1] & 5.4 & 4.0 & [   5.2,   5.2,   6.2] \\ 
49  &  $\text{Na}_{4}\text{Se}\text{Te}$-123 & [ 5.0, 5.0, 7.1] & [ 6, 6, 4] & [ 20, 20, 14] & [ 5.0, 5.0, 7.1] & [ 3, 3, 2] & 6.5 & 4.0 & [   2.5,   2.5,   2.1] \\ 
50  &  $\text{Ru}_{8}\text{Si}_{4}\text{Zr}_{4}$-225 & [ 4.4, 4.4, 4.4] & [ 7, 7, 7] & [ 24, 24, 24] & [ 6.2, 6.2, 6.2] & [ 2, 2, 2] & 6.0 & 4.0 & [  12.0,  12.0,  12.0] \\ 
51  &  $\text{Mg}_{8}\text{Sc}_{16}\text{Se}_{32}$-227 & [ 7.9, 7.9, 7.9] & [ 4, 4, 4] & [ 14, 14, 14] & [11.2,11.2,11.2] & [ 1, 1, 1] & 5.5 & 3.0 & [   5.9,   5.9,   5.8] \\ 
52  &  $\text{As}\text{In}_{2}\text{P}$-115 & [ 4.3, 4.3, 6.1] & [ 7, 7, 5] & [ 24, 24, 16] & [ 4.3, 4.3, 6.1] & [ 3, 3, 2] & 6.0 & 4.0 & [  40.0,  40.0,  36.6] \\ 
53  &  $\text{Cs}_{2}\text{Na}_{2}\text{Se}_{2}$-129 & [ 5.0, 5.0, 8.2] & [ 6, 6, 4] & [ 20, 20, 12] & [ 5.0, 5.0, 8.2] & [ 3, 3, 2] & 6.5 & 4.0 & [   1.0,   1.0,   0.6] \\ 
54  &  $\text{Co}_{4}\text{Sb}_{4}\text{Ti}_{4}$-216 & [ 4.2, 4.2, 4.2] & [ 7, 7, 7] & [ 24, 24, 24] & [ 5.9, 5.9, 5.9] & [ 3, 3, 3] & 6.5 & 4.0 & [  23.0,  23.0,  23.0] \\ 
55  &  $\text{Cs}_{8}\text{N}_{16}\text{V}_{8}$-227 & [ 5.9, 5.9, 5.9] & [ 5, 5, 5] & [ 18, 18, 18] & [ 8.4, 8.4, 8.4] & [ 2, 2, 2] & 6.5 & 4.0 & [   9.9,   9.9,   9.9] \\ 
56  &  $\text{Cd}_{8}\text{Se}_{32}\text{Y}_{16}$-227 & [ 8.3, 8.3, 8.3] & [ 4, 4, 4] & [ 12, 12, 12] & [11.8,11.8,11.8] & [ 1, 1, 1] & 5.5 & 3.0 & [   3.3,   3.3,   3.3] \\ 
57  &  $\text{Ni}_{2}\text{P}_{8}\text{Si}_{6}$-121 & [ 6.4, 6.4, 6.4] & [ 5, 5, 5] & [ 16, 16, 16] & [ 5.2, 5.2,10.5] & [ 3, 3, 1] & 5.0 & 4.0 & [  35.7,  35.7,  33.5] \\ 
58  &  $\text{Ga}_{5}\text{P}_{3}\text{S}_{3}$-25 & [ 3.8,11.4, 5.5] & [ 8, 3, 5] & [ 26, 10, 18] & [ 3.8,11.4, 5.5] & [ 4, 1, 3] & 5.7 & 3.0 & [  26.5,   9.3,  13.5] \\ 
59  &  $\text{Ca}_{4}\text{Li}_{4}\text{P}_{4}$-62 & [ 7.1, 4.2, 7.8] & [ 4, 7, 4] & [ 14, 24, 14] & [ 7.1, 4.2, 7.8] & [ 2, 4, 2] & 5.5 & 3.0 & [   4.7,   5.6,   3.2] \\ 
60  &  $\text{K}_{4}\text{Rb}_{4}\text{S}_{4}$-62 & [ 8.4, 5.0, 9.6] & [ 4, 6, 3] & [ 12, 20, 10] & [ 8.4, 5.0, 9.6] & [ 2, 3, 2] & 6.0 & 3.5 & [   0.5,   0.6,   0.6] \\ 
61  &  $\text{Cu}_{3}\text{S}_{4}\text{Ta}$-215 & [ 5.6, 5.6, 5.6] & [ 5, 5, 5] & [ 18, 18, 18] & [ 5.6, 5.6, 5.6] & [ 3, 3, 3] & 6.5 & 4.0 & [   4.2,   4.2,   4.2] \\ 
62  &  $\text{Mg}_{8}\text{Se}_{32}\text{Y}_{16}$-227 & [ 8.3, 8.3, 8.3] & [ 4, 4, 4] & [ 12, 12, 12] & [11.7,11.7,11.7] & [ 1, 1, 1] & 5.5 & 3.0 & [   5.0,   5.0,   5.0] \\ 
63  &  $\text{Cu}_{3}\text{Ta}\text{Te}_{4}$-215 & [ 6.0, 6.0, 6.0] & [ 5, 5, 5] & [ 18, 18, 18] & [ 6.0, 6.0, 6.0] & [ 2, 2, 2] & 6.0 & 3.0 & [   5.4,   5.4,   5.4] \\ 
64  &  $\text{Ca}_{2}\text{Rb}_{2}\text{Sb}_{2}$-129 & [ 5.4, 5.4, 8.6] & [ 6, 6, 4] & [ 20, 20, 12] & [ 5.4, 5.4, 8.6] & [ 3, 3, 2] & 6.5 & 4.0 & [   1.6,   1.6,   0.7] \\ 
65  &  $\text{Cu}\text{Ga}_{5}\text{Se}_{8}$-111 & [ 5.5, 5.5,11.3] & [ 5, 5, 3] & [ 18, 18, 10] & [ 5.5, 5.5,11.3] & [ 3, 3, 1] & 5.6 & 4.0 & [   6.5,   6.5,   4.4] \\ 
66  &  $\text{Li}_{2}\text{N}_{2}\text{Sr}_{2}$-131 & [ 3.9, 3.9, 7.1] & [ 8, 8, 4] & [ 26, 26, 14] & [ 3.9, 3.9, 7.1] & [ 4, 4, 2] & 6.5 & 4.0 & [  13.7,  13.7,   7.2] \\ 
67  &  $\text{Ag}\text{In}_{5}\text{Se}_{8}$-111 & [ 5.9, 5.9,12.1] & [ 5, 5, 2] & [ 18, 18,  8] & [ 5.9, 5.9,12.1] & [ 3, 3, 1] & 6.0 & 4.0 & [   2.7,   2.7,   1.2] \\ 
68  &  $\text{Mg}_{8}\text{S}_{32}\text{Y}_{16}$-227 & [ 7.9, 7.9, 7.9] & [ 4, 4, 4] & [ 14, 14, 14] & [11.2,11.2,11.2] & [ 1, 1, 1] & 5.5 & 3.0 & [   7.4,   7.4,   7.4] \\ 
69  &  $\text{Al}_{4}\text{Cd}_{2}\text{S}_{8}$-82 & [ 6.5, 6.5, 6.5] & [ 5, 5, 5] & [ 16, 16, 16] & [ 5.6, 5.6,10.3] & [ 3, 3, 1] & 5.1 & 3.0 & [   6.5,   6.5,   7.1] \\ 
70  &  $\text{Ge}_{2}\text{S}_{8}\text{Zn}_{4}$-121 & [ 6.4, 6.4, 6.4] & [ 5, 5, 5] & [ 16, 16, 16] & [ 5.5, 5.5,10.1] & [ 3, 3, 1] & 5.0 & 4.0 & [   9.9,   9.9,  16.1] \\ 
71  &  $\text{Al}_{20}\text{Cu}_{4}\text{Se}_{32}$-216 & [ 7.5, 7.5, 7.5] & [ 4, 4, 4] & [ 14, 14, 14] & [10.6,10.6,10.6] & [ 1, 1, 1] & 5.0 & 3.0 & [   5.3,   5.3,   5.3] \\ 
72  &  $\text{Bi}_{4}\text{Li}_{8}\text{Tl}_{4}$-225 & [ 5.1, 5.1, 5.1] & [ 6, 6, 6] & [ 20, 20, 20] & [ 7.2, 7.2, 7.2] & [ 2, 2, 2] & 6.5 & 4.0 & [   2.4,   2.4,   2.4] \\ 
73  &  $\text{Li}_{4}\text{Mg}_{4}\text{Sb}_{4}$-216 & [ 4.7, 4.7, 4.7] & [ 6, 6, 6] & [ 22, 22, 22] & [ 6.7, 6.7, 6.7] & [ 2, 2, 2] & 6.5 & 4.0 & [   8.9,   8.9,   8.9] \\ 
74  &  $\text{In}_{6}\text{Na}_{6}\text{Se}_{12}$-166 & [ 4.0, 4.0,21.2] & [ 7, 7, 1] & [ 26, 26,  6] & [ 7.0, 4.0,21.2] & [ 2, 4, 1] & 6.5 & 4.0 & [   1.4,   1.4,   0.6] \\ 
75  &  $\text{Ca}_{4}\text{Li}_{4}\text{Sb}_{4}$-62 & [ 7.6, 4.6, 8.3] & [ 4, 6, 4] & [ 14, 22, 12] & [ 7.6, 4.6, 8.3] & [ 2, 3, 2] & 6.0 & 3.5 & [   3.8,   4.1,   2.8] \\ 
76  &  $\text{Mg}_{3}\text{N}\text{Sb}$-221 & [ 4.4, 4.4, 4.4] & [ 7, 7, 7] & [ 24, 24, 24] & [ 4.4, 4.4, 4.4] & [ 3, 3, 3] & 6.5 & 4.0 & [  11.4,  11.4,  11.4] \\ 
77  &  $\text{Cd}_{4}\text{P}_{4}\text{Sr}_{2}$-164 & [ 4.4, 4.4, 7.3] & [ 7, 7, 4] & [ 24, 24, 14] & [ 7.6, 4.4, 7.3] & [ 2, 3, 2] & 6.5 & 4.0 & [   3.7,   3.6,   2.7] \\ 
78  &  $\text{P}_{4}\text{Sr}_{2}\text{Zn}_{4}$-164 & [ 4.1, 4.1, 7.1] & [ 7, 7, 4] & [ 24, 24, 14] & [ 7.1, 4.1, 7.1] & [ 2, 4, 2] & 6.5 & 4.0 & [   7.7,   7.6,   5.3] \\ 
79  &  $\text{B}\text{Ca}_{3}\text{N}_{3}$-123 & [ 3.6, 3.6, 8.2] & [ 8, 8, 4] & [ 28, 28, 12] & [ 3.6, 3.6, 8.2] & [ 4, 4, 2] & 6.5 & 4.0 & [   9.8,   9.8,   3.4] \\ 
80  &  $\text{Al}_{16}\text{S}_{32}\text{Zn}_{8}$-227 & [ 7.1, 7.1, 7.1] & [ 4, 4, 4] & [ 14, 14, 14] & [10.1,10.1,10.1] & [ 1, 1, 1] & 5.0 & 3.0 & [  10.4,  10.4,  10.4] \\ 
81  &  $\text{K}_{4}\text{Sb}_{4}\text{Zn}_{4}$-194 & [ 4.6, 4.6,10.7] & [ 7, 7, 3] & [ 22, 22, 10] & [ 4.6, 7.9,10.7] & [ 3, 2, 1] & 6.5 & 4.0 & [   4.3,   4.3,   0.8] \\ 
82  &  $\text{Al}_{4}\text{Sr}_{2}\text{Te}_{8}$-97 & [ 6.8, 6.8, 6.8] & [ 4, 4, 4] & [ 16, 16, 16] & [ 6.7, 8.4, 8.4] & [ 2, 2, 2] & 6.5 & 4.0 & [   2.7,   0.7,   0.7] \\ 
83  &  $\text{Ga}_{4}\text{Te}_{8}\text{Zn}_{2}$-82 & [ 7.4, 7.4, 7.4] & [ 4, 4, 4] & [ 14, 14, 14] & [ 6.1, 6.1,12.0] & [ 2, 2, 1] & 6.0 & 4.0 & [   4.6,   4.6,   5.7] \\ 
84  &  $\text{Cu}_{4}\text{Li}_{8}\text{Sb}_{4}$-216 & [ 4.5, 4.5, 4.5] & [ 7, 7, 7] & [ 22, 22, 22] & [ 6.3, 6.3, 6.3] & [ 2, 2, 2] & 6.0 & 3.0 & [   5.2,   5.2,   5.2] \\ 
85  &  $\text{Na}_{2}\text{P}_{2}\text{Zn}_{2}$-129 & [ 4.1, 4.1, 6.9] & [ 7, 7, 4] & [ 26, 26, 14] & [ 4.1, 4.1, 6.9] & [ 4, 4, 2] & 6.5 & 4.0 & [   2.5,   2.5,   1.6] \\ 
86  &  $\text{Ag}\text{In}_{5}\text{Te}_{8}$-111 & [ 6.3, 6.3,12.9] & [ 5, 5, 2] & [ 16, 16,  8] & [ 6.3, 6.3,12.9] & [ 2, 2, 1] & 6.3 & 4.0 & [   1.7,   1.7,   1.0] \\ 
87  &  $\text{Ba}_{2}\text{Ga}_{4}\text{Te}_{8}$-97 & [ 7.0, 7.0, 7.0] & [ 4, 4, 4] & [ 14, 14, 14] & [ 8.6, 8.6, 6.8] & [ 2, 2, 2] & 6.5 & 4.0 & [   0.8,   0.8,   2.9] \\ 
88  &  $\text{As}_{3}\text{Cd}_{4}\text{Cs}$-123 & [ 4.6, 4.6,11.0] & [ 7, 7, 3] & [ 22, 22, 10] & [ 4.6, 4.6,11.0] & [ 3, 3, 1] & 5.4 & 4.0 & [   0.9,   0.9,   0.2] \\ 
89  &  $\text{Li}_{4}\text{N}_{4}\text{Zn}_{4}$-216 & [ 3.5, 3.5, 3.5] & [ 9, 9, 9] & [ 30, 30, 30] & [ 4.9, 4.9, 4.9] & [ 3, 3, 3] & 6.5 & 4.0 & [  14.0,  14.0,  14.0] \\ 
90  &  $\text{Bi}_{4}\text{Li}_{4}\text{Sr}_{4}$-62 & [ 4.9, 8.7, 8.2] & [ 6, 3, 4] & [ 20, 12, 12] & [ 4.9, 8.7, 8.2] & [ 3, 2, 2] & 6.5 & 4.0 & [   3.4,   1.9,   2.6] \\ 
91  &  $\text{As}_{2}\text{Na}_{2}\text{Zn}_{2}$-129 & [ 4.2, 4.2, 7.1] & [ 7, 7, 4] & [ 24, 24, 14] & [ 4.2, 4.2, 7.1] & [ 4, 4, 2] & 6.5 & 4.0 & [   1.8,   1.8,   1.2] \\ 
92  &  $\text{Cs}_{4}\text{K}_{8}\text{Sb}_{4}$-225 & [ 6.2, 6.2, 6.2] & [ 5, 5, 5] & [ 16, 16, 16] & [ 8.8, 8.8, 8.8] & [ 2, 2, 2] & 6.5 & 4.0 & [   0.2,   0.2,   0.2] \\ 
93  &  $\text{As}_{12}\text{Ca}_{6}\text{Ga}_{12}$-166 & [ 4.0, 4.0,25.1] & [ 7, 7, 1] & [ 26, 26,  4] & [ 7.0, 4.0,25.1] & [ 2, 4, 1] & 6.0 & 3.0 & [   6.8,   6.9,   5.4] \\ 
94  &  $\text{Al}\text{Ga}\text{N}_{2}$-115 & [ 3.2, 3.2, 4.5] & [ 9, 9, 7] & [ 32, 32, 22] & [ 3.2, 3.2, 4.5] & [ 5, 5, 3] & 5.5 & 2.5 & [ 125.5, 125.5,  91.0] \\ 
95  &  $\text{C}_{3}\text{Cd}_{3}\text{N}_{6}$-166 & [ 3.6, 3.6,14.6] & [ 8, 8, 2] & [ 28, 28,  8] & [ 3.6, 3.6,14.6] & [ 4, 4, 1] & 5.5 & 3.5 & [   2.7,   2.7,   3.6] \\ 
96  &  $\text{Bi}_{4}\text{Ni}_{4}\text{Sc}_{4}$-216 & [ 4.4, 4.4, 4.4] & [ 7, 7, 7] & [ 24, 24, 24] & [ 6.3, 6.3, 6.3] & [ 2, 2, 2] & 6.5 & 4.0 & [   5.3,   5.3,   5.3] \\ 
97  &  $\text{K}_{2}\text{Mg}_{2}\text{P}_{2}$-129 & [ 4.5, 4.5, 7.6] & [ 7, 7, 4] & [ 22, 22, 14] & [ 4.5, 4.5, 7.6] & [ 3, 3, 2] & 6.5 & 4.0 & [   5.1,   5.1,   1.8] \\ 
98  &  $\text{Cd}\text{In}_{2}\text{Se}_{4}$-111 & [ 5.9, 5.9, 6.1] & [ 5, 5, 5] & [ 18, 18, 16] & [ 5.9, 5.9, 6.1] & [ 3, 3, 2] & 6.0 & 4.0 & [   6.0,   6.0,   4.0] \\ 
99  &  $\text{Li}_{4}\text{Na}_{4}\text{Se}_{4}$-62 & [ 7.2, 4.3, 7.8] & [ 4, 7, 4] & [ 14, 24, 14] & [ 7.2, 4.3, 7.8] & [ 2, 4, 2] & 6.5 & 4.0 & [   2.6,   3.5,   2.4] \\ 
100  &  $\text{In}_{4}\text{Se}_{8}\text{Zn}_{2}$-82 & [ 7.2, 7.2, 7.2] & [ 4, 4, 4] & [ 14, 14, 14] & [ 5.8, 5.8,11.8] & [ 3, 3, 1] & 5.8 & 4.0 & [   5.1,   5.1,   6.4] \\ 
101  &  $\text{Bi}_{2}\text{Ca}_{2}\text{K}_{2}$-129 & [ 5.4, 5.4, 8.5] & [ 6, 6, 4] & [ 18, 18, 12] & [ 5.4, 5.4, 8.5] & [ 3, 3, 2] & 6.5 & 4.0 & [   1.3,   1.3,   0.5] \\ 
102  &  $\text{As}_{4}\text{Mg}_{4}\text{Sr}_{2}$-164 & [ 4.4, 4.4, 7.4] & [ 7, 7, 4] & [ 24, 24, 14] & [ 7.7, 4.4, 7.4] & [ 2, 3, 2] & 6.5 & 4.0 & [   6.5,   6.5,   6.1] \\ 
103  &  $\text{Li}_{2}\text{Rb}_{2}\text{S}_{2}$-129 & [ 4.4, 4.4, 7.4] & [ 7, 7, 4] & [ 24, 24, 14] & [ 4.4, 4.4, 7.4] & [ 3, 3, 2] & 6.5 & 4.0 & [   1.9,   1.9,   1.0] \\ 
104  &  $\text{Ba}_{2}\text{Mg}_{4}\text{P}_{4}$-164 & [ 4.4, 4.4, 7.6] & [ 7, 7, 4] & [ 24, 24, 14] & [ 7.6, 4.4, 7.6] & [ 2, 3, 2] & 6.5 & 4.0 & [   6.2,   6.3,   4.1] \\ 
105  &  $\text{Co}_{4}\text{Nb}_{4}\text{Sn}_{4}$-216 & [ 4.2, 4.2, 4.2] & [ 7, 7, 7] & [ 24, 24, 24] & [ 6.0, 6.0, 6.0] & [ 3, 3, 3] & 6.5 & 4.0 & [  17.6,  17.6,  17.5] \\ 
106  &  $\text{Cd}_{8}\text{S}_{32}\text{Y}_{16}$-227 & [ 8.0, 8.0, 8.0] & [ 4, 4, 4] & [ 14, 14, 14] & [11.3,11.3,11.3] & [ 1, 1, 1] & 5.5 & 4.0 & [   5.0,   5.0,   5.1] \\ 
107  &  $\text{Ba}_{4}\text{Li}_{4}\text{Sb}_{4}$-194 & [ 4.9, 4.9, 9.2] & [ 6, 6, 3] & [ 20, 20, 12] & [ 4.9, 8.5, 9.2] & [ 3, 2, 2] & 6.5 & 4.0 & [   5.1,   5.1,   4.0] \\ 
108  &  $\text{Na}_{2}\text{Rb}_{2}\text{S}_{2}$-129 & [ 4.7, 4.7, 7.7] & [ 6, 6, 4] & [ 22, 22, 14] & [ 4.7, 4.7, 7.7] & [ 3, 3, 2] & 6.5 & 4.0 & [   1.4,   1.4,   0.9] \\ 
109  &  $\text{Na}_{3}\text{P}_{3}\text{Sr}_{3}$-189 & [ 7.7, 4.5, 7.7] & [ 4, 7, 4] & [ 14, 22, 14] & [ 7.7, 4.5, 7.7] & [ 2, 3, 2] & 6.5 & 4.0 & [   1.5,   1.8,   1.5] \\ 
110  &  $\text{As}_{4}\text{Ba}_{2}\text{Mg}_{4}$-164 & [ 4.5, 4.5, 7.8] & [ 7, 7, 4] & [ 22, 22, 14] & [ 7.8, 4.5, 7.8] & [ 2, 3, 2] & 6.5 & 4.0 & [   5.5,   5.5,   4.0] \\ 
111  &  $\text{Cs}_{2}\text{Na}_{2}\text{Te}_{2}$-129 & [ 5.4, 5.4, 8.7] & [ 6, 6, 3] & [ 20, 20, 12] & [ 5.4, 5.4, 8.7] & [ 3, 3, 2] & 6.0 & 3.0 & [   1.1,   1.1,   0.6] \\ 
112  &  $\text{As}_{4}\text{Be}_{4}\text{Na}_{4}$-194 & [ 3.8, 3.8, 9.0] & [ 8, 8, 3] & [ 26, 26, 12] & [ 3.8, 6.6, 9.0] & [ 4, 2, 2] & 6.5 & 4.0 & [  29.2,  29.2,   3.5] \\ 
113  &  $\text{Ba}_{2}\text{Mg}_{4}\text{Sb}_{4}$-164 & [ 4.8, 4.8, 8.2] & [ 6, 6, 4] & [ 22, 22, 12] & [ 8.3, 4.8, 8.2] & [ 2, 3, 2] & 6.5 & 4.0 & [   7.9,   7.9,   7.7] \\ 
114  &  $\text{In}_{6}\text{S}_{12}\text{Tl}_{6}$-166 & [ 3.9, 3.9,22.4] & [ 8, 8, 1] & [ 26, 26,  4] & [ 6.8, 3.9,22.4] & [ 2, 4, 1] & 6.0 & 3.0 & [   1.2,   1.2,   0.7] \\ 
115  &  $\text{As}_{4}\text{Ca}_{4}\text{Na}_{4}$-216 & [ 4.9, 4.9, 4.9] & [ 6, 6, 6] & [ 20, 20, 20] & [ 7.0, 7.0, 7.0] & [ 2, 2, 2] & 6.5 & 4.0 & [   7.3,   7.3,   7.3] \\ 
116  &  $\text{As}_{2}\text{Mg}_{2}\text{Na}_{2}$-129 & [ 4.4, 4.4, 7.2] & [ 7, 7, 4] & [ 24, 24, 14] & [ 4.4, 4.4, 7.2] & [ 3, 3, 2] & 6.5 & 4.0 & [   3.1,   3.1,   1.3] \\ 
117  &  $\text{Bi}_{4}\text{In}_{4}\text{Li}_{8}$-225 & [ 5.0, 5.0, 5.0] & [ 6, 6, 6] & [ 20, 20, 20] & [ 7.1, 7.1, 7.1] & [ 2, 2, 2] & 6.5 & 4.0 & [   5.0,   5.0,   5.0] \\ 
118  &  $\text{C}_{3}\text{Ca}_{3}\text{N}_{6}$-166 & [ 3.7, 3.7,14.8] & [ 8, 8, 2] & [ 28, 28,  8] & [ 3.7, 3.7,14.8] & [ 4, 4, 1] & 5.5 & 3.5 & [  15.8,  15.8,  13.9] \\ 
119  &  $\text{As}_{4}\text{Co}_{4}\text{Se}_{4}$-198 & [ 5.8, 5.8, 5.8] & [ 5, 5, 5] & [ 18, 18, 18] & [ 5.8, 5.8, 5.8] & [ 3, 3, 3] & 6.0 & 2.5 & [  19.7,  19.7,  19.7] \\ 
120  &  $\text{In}_{16}\text{Mg}_{8}\text{S}_{32}$-227 & [ 7.7, 7.7, 7.7] & [ 4, 4, 4] & [ 14, 14, 14] & [10.9,10.9,10.9] & [ 1, 1, 1] & 5.4 & 4.0 & [   5.2,   5.2,   5.2] \\ 
121  &  $\text{Al}\text{Ga}\text{P}_{2}$-115 & [ 3.9, 3.9, 5.5] & [ 8, 8, 5] & [ 26, 26, 18] & [ 3.9, 3.9, 5.5] & [ 4, 4, 3] & 6.5 & 4.0 & [  46.3,  46.3,  35.7] \\ 
122  &  $\text{Pb}_{2}\text{S}_{4}\text{Sn}_{2}$-26 & [ 4.4, 4.1,11.7] & [ 7, 7, 3] & [ 24, 24, 10] & [ 4.4, 4.1,11.7] & [ 3, 4, 1] & 5.8 & 3.0 & [   1.1,   1.5,   0.5] \\ 
123  &  $\text{In}_{2}\text{P}_{2}\text{S}_{8}$-82 & [ 6.2, 6.2, 6.2] & [ 5, 5, 5] & [ 16, 16, 16] & [ 5.9, 5.9, 9.2] & [ 3, 3, 2] & 6.5 & 4.0 & [   3.2,   3.2,   3.3] \\ 
124  &  $\text{Li}_{4}\text{Mg}_{4}\text{P}_{4}$-216 & [ 4.2, 4.2, 4.2] & [ 7, 7, 7] & [ 24, 24, 24] & [ 6.0, 6.0, 6.0] & [ 2, 2, 2] & 6.0 & 4.0 & [  10.1,  10.1,  10.1] \\ 
125  &  $\text{As}_{3}\text{Rb}\text{Zn}_{4}$-123 & [ 4.2, 4.2,10.5] & [ 7, 7, 3] & [ 24, 24, 10] & [ 4.2, 4.2,10.5] & [ 4, 4, 1] & 5.0 & 4.0 & [   1.4,   1.4,   0.4] \\ 
126  &  $\text{Cu}_{3}\text{Nb}\text{S}_{4}$-215 & [ 5.5, 5.5, 5.5] & [ 5, 5, 5] & [ 18, 18, 18] & [ 5.5, 5.5, 5.5] & [ 3, 3, 3] & 6.5 & 4.0 & [   3.6,   3.6,   3.6] \\ 
127  &  $\text{Be}_{4}\text{Li}_{4}\text{Sb}_{4}$-186 & [ 4.2, 4.2, 6.7] & [ 7, 7, 4] & [ 24, 24, 16] & [ 7.2, 4.2, 6.7] & [ 2, 4, 2] & 5.0 & 2.6 & [  14.7,  14.7,  19.3] \\ 
128  &  $\text{Bi}_{4}\text{Ni}_{4}\text{Y}_{4}$-216 & [ 4.6, 4.6, 4.6] & [ 7, 7, 7] & [ 22, 22, 22] & [ 6.5, 6.5, 6.5] & [ 2, 2, 2] & 6.5 & 4.0 & [   6.3,   6.3,   6.3] \\ 
129  &  $\text{Li}_{3}\text{Nb}\text{S}_{4}$-215 & [ 6.1, 6.1, 6.1] & [ 5, 5, 5] & [ 16, 16, 16] & [ 6.1, 6.1, 6.1] & [ 2, 2, 2] & 6.5 & 4.0 & [   2.0,   2.0,   2.0] \\ 
130  &  $\text{Ni}_{4}\text{Sb}_{4}\text{Sc}_{4}$-216 & [ 4.3, 4.3, 4.3] & [ 7, 7, 7] & [ 24, 24, 24] & [ 6.1, 6.1, 6.1] & [ 2, 2, 2] & 6.0 & 3.0 & [  13.8,  13.8,  13.8] \\ 
131  &  $\text{Ag}_{2}\text{K}_{2}\text{Se}_{2}$-129 & [ 4.6, 4.6, 7.8] & [ 7, 7, 4] & [ 22, 22, 14] & [ 4.6, 4.6, 7.8] & [ 3, 3, 2] & 6.5 & 4.0 & [   0.4,   0.4,   0.3] \\ 
132  &  $\text{Cd}_{8}\text{In}_{16}\text{S}_{32}$-227 & [ 7.8, 7.8, 7.8] & [ 4, 4, 4] & [ 14, 14, 14] & [11.0,11.0,11.0] & [ 1, 1, 1] & 5.5 & 3.0 & [   4.5,   4.5,   4.6] \\ 
133  &  $\text{Al}_{20}\text{Cu}_{4}\text{S}_{32}$-216 & [ 7.1, 7.1, 7.1] & [ 4, 4, 4] & [ 14, 14, 14] & [10.0,10.0,10.0] & [ 1, 1, 1] & 5.0 & 3.0 & [   7.0,   7.0,   7.1] \\ 
134  &  $\text{Be}_{2}\text{Li}_{2}\text{P}_{2}$-129 & [ 3.6, 3.6, 6.0] & [ 8, 8, 5] & [ 28, 28, 18] & [ 3.6, 3.6, 6.0] & [ 4, 4, 2] & 6.0 & 4.0 & [  10.0,  10.0,   4.1] \\ 
135  &  $\text{Cd}\text{S}_{4}\text{Zn}_{3}$-215 & [ 5.6, 5.6, 5.6] & [ 5, 5, 5] & [ 18, 18, 18] & [ 5.6, 5.6, 5.6] & [ 3, 3, 3] & 6.5 & 4.0 & [  13.6,  13.6,  13.6] \\ 
136  &  $\text{In}_{4}\text{Mg}_{2}\text{Te}_{8}$-82 & [ 7.8, 7.8, 7.8] & [ 4, 4, 4] & [ 14, 14, 14] & [ 6.4, 6.4,12.7] & [ 2, 2, 1] & 6.3 & 4.0 & [   2.8,   2.8,   4.0] \\ 
137  &  $\text{As}_{4}\text{Ca}_{2}\text{Mg}_{4}$-164 & [ 4.4, 4.4, 7.1] & [ 7, 7, 4] & [ 24, 24, 14] & [ 7.6, 4.4, 7.1] & [ 2, 3, 2] & 6.5 & 3.0 & [   5.5,   5.6,   4.4] \\ 
138  &  $\text{As}_{4}\text{Be}_{4}\text{Mg}_{2}$-164 & [ 3.8, 3.8, 6.5] & [ 8, 8, 5] & [ 26, 26, 16] & [ 6.6, 3.8, 6.5] & [ 2, 4, 2] & 6.0 & 2.5 & [  11.0,  11.2,   6.8] \\ 
139  &  $\text{Na}_{8}\text{Rb}_{4}\text{Sb}_{4}$-225 & [ 5.6, 5.6, 5.6] & [ 5, 5, 5] & [ 18, 18, 18] & [ 7.9, 7.9, 7.9] & [ 2, 2, 2] & 6.5 & 4.0 & [   0.4,   0.4,   0.4] \\ 
140  &  $\text{In}_{4}\text{Mg}_{2}\text{Te}_{8}$-121 & [ 7.8, 7.8, 7.8] & [ 4, 4, 4] & [ 14, 14, 14] & [ 6.4, 6.4,12.8] & [ 2, 2, 1] & 6.3 & 4.0 & [   5.7,   5.7,   4.4] \\ 
141  &  $\text{In}_{4}\text{Se}_{8}\text{Zn}_{2}$-121 & [ 7.1, 7.1, 7.1] & [ 4, 4, 4] & [ 14, 14, 14] & [ 5.9, 5.9,11.5] & [ 3, 3, 1] & 5.7 & 4.0 & [   6.4,   6.4,   6.9] \\ 
142  &  $\text{Li}_{4}\text{P}_{4}\text{Zn}_{4}$-216 & [ 4.1, 4.1, 4.1] & [ 7, 7, 7] & [ 26, 26, 26] & [ 5.7, 5.7, 5.7] & [ 3, 3, 3] & 6.5 & 4.0 & [   7.2,   7.2,   7.3] \\ 
143  &  $\text{K}_{4}\text{P}_{4}\text{Zn}_{4}$-194 & [ 4.1, 4.1,10.3] & [ 7, 7, 3] & [ 24, 24, 10] & [ 4.1, 7.1,10.3] & [ 4, 2, 1] & 5.2 & 4.0 & [   9.8,   9.8,   3.0] \\ 
144  &  $\text{Ca}_{2}\text{S}_{4}\text{Sn}$-65 & [ 7.0, 3.9, 7.0] & [ 4, 8, 4] & [ 14, 26, 14] & [ 7.0, 3.9, 7.0] & [ 2, 4, 2] & 6.5 & 4.0 & [   2.0,   3.8,   2.0] \\ 
145  &  $\text{N}_{8}\text{S}_{4}\text{Zr}_{8}$-194 & [ 3.6, 3.6,12.9] & [ 8, 8, 2] & [ 28, 28,  8] & [ 6.3, 3.6,12.9] & [ 2, 4, 1] & 5.5 & 2.5 & [  11.8,  10.6,   6.7] \\ 
146  &  $\text{Mg}\text{S}_{2}\text{Zn}$-115 & [ 3.9, 3.9, 5.6] & [ 8, 8, 5] & [ 26, 26, 18] & [ 3.9, 3.9, 5.6] & [ 4, 4, 3] & 6.5 & 4.0 & [  16.6,  16.6,  12.2] \\ 
147  &  $\text{As}_{4}\text{Be}_{4}\text{Ca}_{2}$-164 & [ 3.9, 3.9, 6.8] & [ 8, 8, 4] & [ 26, 26, 16] & [ 6.7, 3.9, 6.8] & [ 2, 4, 2] & 6.5 & 4.0 & [  33.3,  34.2,  14.9] \\ 
148  &  $\text{Mg}_{2}\text{S}\text{Se}$-115 & [ 4.1, 4.1, 5.8] & [ 7, 7, 5] & [ 24, 24, 18] & [ 4.1, 4.1, 5.8] & [ 4, 4, 3] & 6.5 & 4.0 & [  17.0,  17.5,  13.5] \\ 
149  &  $\text{Bi}_{6}\text{Na}_{6}\text{Se}_{12}$-166 & [ 4.2, 4.2,21.3] & [ 7, 7, 1] & [ 24, 24,  6] & [ 7.3, 4.2,21.3] & [ 2, 4, 1] & 6.5 & 4.0 & [   1.1,   1.1,   0.5] \\ 
150  &  $\text{As}_{2}\text{K}_{2}\text{Mg}_{2}$-129 & [ 4.6, 4.6, 7.9] & [ 7, 7, 4] & [ 22, 22, 14] & [ 4.6, 4.6, 7.9] & [ 3, 3, 2] & 6.5 & 4.0 & [   4.8,   4.8,   1.5] \\ 
151  &  $\text{Cd}\text{S}_{2}\text{Zn}$-115 & [ 4.0, 4.0, 5.7] & [ 7, 7, 5] & [ 26, 26, 18] & [ 4.0, 4.0, 5.7] & [ 4, 4, 3] & 6.5 & 4.0 & [  14.7,  14.7,  12.9] \\ 
152  &  $\text{Li}_{4}\text{Nb}_{4}\text{Se}_{8}$-194 & [ 3.5, 3.5,13.6] & [ 8, 8, 2] & [ 28, 28,  8] & [ 3.5, 6.1,13.6] & [ 4, 2, 1] & 6.0 & 4.0 & [  10.8,  11.0,   2.0] \\ 
153  &  $\text{Li}_{8}\text{Na}_{4}\text{Sb}_{4}$-225 & [ 4.8, 4.8, 4.8] & [ 6, 6, 6] & [ 22, 22, 22] & [ 6.8, 6.8, 6.8] & [ 2, 2, 2] & 6.5 & 4.0 & [   6.1,   6.1,   6.1] \\ 
154  &  $\text{Ga}_{2}\text{S}_{4}\text{Zn}$-111 & [ 5.2, 5.2, 5.4] & [ 6, 6, 6] & [ 20, 20, 18] & [ 5.2, 5.2, 5.4] & [ 3, 3, 3] & 6.5 & 4.0 & [  13.3,  13.3,   9.6] \\ 
155  &  $\text{Ba}_{3}\text{Bi}_{3}\text{Na}_{3}$-189 & [ 8.7, 5.2, 8.7] & [ 3, 6, 3] & [ 12, 20, 12] & [ 8.7, 5.2, 8.7] & [ 2, 3, 2] & 6.0 & 3.0 & [   1.6,   2.4,   1.6] \\ 
156  &  $\text{N}\text{Sb}\text{Sr}_{3}$-221 & [ 5.2, 5.2, 5.2] & [ 6, 6, 6] & [ 20, 20, 20] & [ 5.2, 5.2, 5.2] & [ 3, 3, 3] & 6.5 & 4.0 & [   3.1,   3.1,   3.1] \\ 
157  &  $\text{Cd}_{8}\text{S}_{32}\text{Sc}_{16}$-227 & [ 7.7, 7.7, 7.7] & [ 4, 4, 4] & [ 14, 14, 14] & [10.8,10.8,10.8] & [ 1, 1, 1] & 5.0 & 3.0 & [   5.5,   5.5,   5.5] \\ 
158  &  $\text{N}_{4}\text{Se}_{2}\text{Zr}_{4}$-164 & [ 3.7, 3.7, 6.7] & [ 8, 8, 4] & [ 28, 28, 16] & [ 6.3, 3.7, 6.7] & [ 2, 4, 2] & 6.0 & 2.5 & [   8.7,   8.6,   4.5] \\ 
159  &  $\text{Cu}_{3}\text{Te}_{4}\text{V}$-215 & [ 6.0, 6.0, 6.0] & [ 5, 5, 5] & [ 18, 18, 18] & [ 6.0, 6.0, 6.0] & [ 3, 3, 3] & 6.5 & 4.0 & [   4.9,   4.9,   4.9] \\ 
160  &  $\text{As}_{2}\text{Be}_{2}\text{Li}_{2}$-129 & [ 3.7, 3.7, 6.2] & [ 8, 8, 5] & [ 28, 28, 16] & [ 3.7, 3.7, 6.2] & [ 4, 4, 2] & 6.2 & 4.0 & [  10.4,  10.4,   3.9] \\ 
161  &  $\text{As}_{4}\text{Li}_{4}\text{Zn}_{4}$-216 & [ 4.2, 4.2, 4.2] & [ 7, 7, 7] & [ 24, 24, 24] & [ 6.0, 6.0, 6.0] & [ 3, 3, 3] & 6.5 & 4.0 & [   5.4,   5.4,   5.4] \\ 
162  &  $\text{Ca}_{2}\text{P}_{4}\text{Zn}_{4}$-164 & [ 4.0, 4.0, 6.8] & [ 7, 7, 4] & [ 26, 26, 16] & [ 7.0, 4.0, 6.8] & [ 2, 4, 2] & 6.5 & 4.0 & [   5.6,   5.6,   4.0] \\ 
163  &  $\text{Na}_{4}\text{P}_{4}\text{Sn}_{4}$-186 & [ 3.9, 3.9,11.7] & [ 8, 8, 3] & [ 26, 26, 10] & [ 3.9, 6.8,11.7] & [ 4, 2, 1] & 5.3 & 3.0 & [   4.0,   3.9,   1.1] \\ 
164  &  $\text{Ge}\text{Mg}_{4}\text{Si}$-123 & [ 4.5, 4.5, 6.4] & [ 7, 7, 5] & [ 22, 22, 16] & [ 4.5, 4.5, 6.4] & [ 3, 3, 2] & 6.0 & 3.0 & [   6.8,   6.8,   6.3] \\ 
165  &  $\text{Ge}_{6}\text{Na}_{2}\text{P}_{6}$-26 & [ 8.5, 3.7,10.5] & [ 4, 8, 3] & [ 12, 28, 10] & [ 8.5, 3.7,10.5] & [ 2, 4, 1] & 5.2 & 3.0 & [   0.5,   6.6,   1.3] \\ 
166  &  $\text{As}_{4}\text{Cd}_{4}\text{Li}_{4}$-216 & [ 4.5, 4.5, 4.5] & [ 7, 7, 7] & [ 22, 22, 22] & [ 6.3, 6.3, 6.3] & [ 2, 2, 2] & 6.0 & 4.0 & [   5.7,   5.7,   5.7] \\ 
167  &  $\text{C}_{3}\text{Mg}_{3}\text{N}_{6}$-166 & [ 3.3, 3.3,14.2] & [ 9, 9, 2] & [ 30, 30,  8] & [ 3.3, 3.3,14.2] & [ 5, 5, 1] & 5.5 & 3.5 & [  19.8,  19.8,  22.8] \\ 
168  &  $\text{Ca}_{3}\text{Pb}\text{Se}_{4}$-47 & [ 4.3, 6.0, 8.6] & [ 7, 5, 4] & [ 24, 18, 12] & [ 4.3, 6.0, 8.6] & [ 4, 2, 2] & 6.0 & 4.0 & [   3.9,   4.6,   3.0] \\ 
169  &  $\text{Ba}_{4}\text{Li}_{4}\text{P}_{4}$-194 & [ 4.5, 4.5, 8.7] & [ 7, 7, 3] & [ 22, 22, 12] & [ 4.5, 7.8, 8.7] & [ 3, 2, 2] & 6.5 & 4.0 & [   2.9,   2.9,   2.3] \\ 
170  &  $\text{As}_{2}\text{Ba}_{2}\text{Li}_{2}$-187 & [ 4.6, 4.5, 4.5] & [ 7, 7, 7] & [ 22, 22, 22] & [ 4.6, 4.5, 7.9] & [ 3, 3, 2] & 6.5 & 4.0 & [   3.1,   3.2,   3.2] \\ 
171  &  $\text{N}_{4}\text{S}_{2}\text{Zr}_{4}$-164 & [ 3.6, 3.6, 6.5] & [ 8, 8, 5] & [ 28, 28, 16] & [ 6.3, 3.6, 6.5] & [ 2, 4, 2] & 5.5 & 2.5 & [  10.2,  10.1,   4.6] \\ 
172  &  $\text{Ga}_{4}\text{S}_{8}\text{Zn}_{2}$-82 & [ 6.5, 6.5, 6.5] & [ 5, 5, 5] & [ 16, 16, 16] & [ 5.4, 5.4,10.5] & [ 3, 3, 1] & 5.2 & 4.0 & [  12.6,  12.6,  17.0] \\ 
173  &  $\text{Ca}\text{Pb}\text{Se}_{2}$-123 & [ 4.3, 4.3, 6.1] & [ 7, 7, 5] & [ 24, 24, 16] & [ 4.3, 4.3, 6.1] & [ 3, 3, 2] & 6.0 & 4.0 & [   3.9,   3.9,   3.4] \\ 
174  &  $\text{Al}_{3}\text{Ga}\text{N}_{4}$-215 & [ 4.4, 4.4, 4.4] & [ 7, 7, 7] & [ 24, 24, 24] & [ 4.4, 4.4, 4.4] & [ 3, 3, 3] & 6.5 & 4.0 & [  77.7,  77.7,  77.7] \\ 
175  &  $\text{Pb}\text{S}_{4}\text{Sr}_{3}$-47 & [ 4.3, 6.0, 8.6] & [ 7, 5, 4] & [ 24, 18, 12] & [ 4.3, 6.0, 8.6] & [ 4, 2, 2] & 6.0 & 4.0 & [   2.6,   3.1,   1.7] \\ 
176  &  $\text{K}_{4}\text{Na}_{4}\text{Se}_{4}$-62 & [ 8.1, 4.8, 8.8] & [ 4, 6, 3] & [ 12, 22, 12] & [ 8.1, 4.8, 8.8] & [ 2, 3, 2] & 6.0 & 3.5 & [   1.2,   1.5,   1.2] \\ 
177  &  $\text{As}_{2}\text{Cd}_{2}\text{K}_{2}$-129 & [ 4.6, 4.6, 8.0] & [ 6, 6, 4] & [ 22, 22, 12] & [ 4.6, 4.6, 8.0] & [ 3, 3, 2] & 6.5 & 4.0 & [   1.6,   1.6,   0.7] \\ 
178  &  $\text{Al}_{4}\text{Ba}_{2}\text{Te}_{8}$-97 & [ 7.0, 7.0, 7.0] & [ 4, 4, 4] & [ 14, 14, 14] & [ 6.7, 8.7, 8.7] & [ 2, 2, 2] & 6.5 & 4.0 & [   3.3,   0.9,   0.8] \\ 
179  &  $\text{Cu}_{3}\text{Se}_{4}\text{V}$-215 & [ 5.7, 5.7, 5.7] & [ 5, 5, 5] & [ 18, 18, 18] & [ 5.7, 5.7, 5.7] & [ 3, 3, 3] & 6.5 & 4.0 & [   5.0,   5.0,   5.0] \\ 
180  &  $\text{Mg}_{4}\text{N}_{4}\text{Sr}_{2}$-164 & [ 3.6, 3.6, 6.4] & [ 8, 8, 5] & [ 28, 28, 16] & [ 6.3, 3.6, 6.4] & [ 2, 4, 2] & 5.5 & 2.5 & [   8.8,   8.8,   5.6] \\ 
181  &  $\text{Li}_{4}\text{P}_{4}\text{Sr}_{4}$-194 & [ 4.4, 4.4, 8.1] & [ 7, 7, 4] & [ 24, 24, 12] & [ 4.4, 7.6, 8.1] & [ 3, 2, 2] & 6.5 & 4.0 & [   3.2,   3.2,   2.9] \\ 
182  &  $\text{Cd}\text{Mg}\text{Te}_{2}$-115 & [ 4.7, 4.7, 6.5] & [ 6, 6, 5] & [ 22, 22, 16] & [ 4.7, 4.7, 6.5] & [ 3, 3, 2] & 6.5 & 4.0 & [   6.4,   6.4,   5.0] \\ 
183  &  $\text{Bi}_{4}\text{K}_{4}\text{Na}_{8}$-225 & [ 5.6, 5.6, 5.6] & [ 5, 5, 5] & [ 18, 18, 18] & [ 8.0, 8.0, 8.0] & [ 2, 2, 2] & 6.5 & 4.0 & [   0.8,   0.8,   0.8] \\ 
184  &  $\text{S}\text{Se}\text{Zn}_{2}$-115 & [ 4.0, 4.0, 5.6] & [ 8, 8, 5] & [ 26, 26, 18] & [ 4.0, 4.0, 5.6] & [ 4, 4, 3] & 6.5 & 4.0 & [  15.1,  15.1,  12.4] \\ 
185  &  $\text{P}_{2}\text{Sn}\text{Zn}$-115 & [ 4.0, 4.0, 5.7] & [ 7, 7, 5] & [ 26, 26, 18] & [ 4.0, 4.0, 5.7] & [ 4, 4, 3] & 6.5 & 4.0 & [  42.6,  41.5,  34.6] \\ 
186  &  $\text{K}_{4}\text{Na}_{4}\text{Te}_{4}$-62 & [ 8.6, 5.1, 9.4] & [ 3, 6, 3] & [ 12, 20, 12] & [ 8.6, 5.1, 9.4] & [ 2, 3, 2] & 6.0 & 3.5 & [   0.8,   1.0,   0.8] \\ 
187  &  $\text{Cu}_{3}\text{S}_{4}\text{V}$-215 & [ 5.4, 5.4, 5.4] & [ 6, 6, 6] & [ 18, 18, 18] & [ 5.4, 5.4, 5.4] & [ 3, 3, 3] & 6.5 & 4.0 & [   6.2,   6.2,   6.1] \\ 
188  &  $\text{As}_{4}\text{Li}_{4}\text{Sr}_{4}$-62 & [ 4.5, 8.2, 7.6] & [ 7, 4, 4] & [ 22, 12, 14] & [ 4.5, 8.2, 7.6] & [ 3, 2, 2] & 6.0 & 3.0 & [   3.5,   1.8,   2.3] \\ 
189  &  $\text{Cd}_{2}\text{Ga}_{4}\text{S}_{8}$-82 & [ 6.5, 6.5, 6.5] & [ 5, 5, 5] & [ 16, 16, 16] & [ 5.6, 5.6,10.4] & [ 3, 3, 1] & 5.1 & 4.0 & [   7.9,   7.9,   8.8] \\ 
190  &  $\text{Bi}_{4}\text{Mg}_{4}\text{Sr}_{2}$-164 & [ 4.9, 4.9, 8.0] & [ 6, 6, 4] & [ 22, 22, 14] & [ 8.4, 4.9, 8.0] & [ 2, 3, 2] & 6.5 & 4.0 & [   4.8,   4.8,   4.7] \\ 
191  &  $\text{Li}_{9}\text{N}\text{S}_{3}$-221 & [ 5.5, 5.5, 5.5] & [ 5, 5, 5] & [ 18, 18, 18] & [ 5.5, 5.5, 5.5] & [ 3, 3, 3] & 6.5 & 4.0 & [   3.0,   3.0,   3.0] \\ 
192  &  $\text{Cd}_{2}\text{Se}\text{Te}$-115 & [ 4.5, 4.5, 6.4] & [ 7, 7, 5] & [ 22, 22, 16] & [ 4.5, 4.5, 6.4] & [ 3, 3, 2] & 6.0 & 4.0 & [   4.5,   4.4,   4.1] \\ 
193  &  $\text{In}_{4}\text{Te}_{8}\text{Zn}_{2}$-82 & [ 7.7, 7.7, 7.7] & [ 4, 4, 4] & [ 14, 14, 14] & [ 6.3, 6.3,12.6] & [ 2, 2, 1] & 6.2 & 4.0 & [   3.7,   3.7,   4.3] \\ 
194  &  $\text{As}_{2}\text{Ga}\text{In}$-115 & [ 4.2, 4.2, 6.0] & [ 7, 7, 5] & [ 24, 24, 18] & [ 4.2, 4.2, 6.0] & [ 4, 4, 2] & 6.0 & 4.0 & [  16.6,  16.8,  15.3] \\ 
195  &  $\text{Bi}_{4}\text{Co}_{4}\text{Zr}_{4}$-216 & [ 4.4, 4.4, 4.4] & [ 7, 7, 7] & [ 24, 24, 24] & [ 6.2, 6.2, 6.2] & [ 2, 2, 2] & 6.5 & 4.0 & [  14.2,  14.2,  14.2] \\ 
196  &  $\text{Li}_{4}\text{Mg}_{4}\text{N}_{4}$-216 & [ 3.5, 3.5, 3.5] & [ 8, 8, 8] & [ 28, 28, 28] & [ 5.0, 5.0, 5.0] & [ 3, 3, 3] & 6.5 & 4.0 & [  18.1,  18.1,  18.1] \\ 
197  &  $\text{Al}\text{Ga}_{3}\text{N}_{4}$-215 & [ 4.5, 4.5, 4.5] & [ 7, 7, 7] & [ 22, 22, 22] & [ 4.5, 4.5, 4.5] & [ 3, 3, 3] & 6.5 & 4.0 & [  66.6,  66.6,  66.6] \\ 
198  &  $\text{Al}_{4}\text{Se}_{8}\text{Zn}_{2}$-82 & [ 6.8, 6.8, 6.8] & [ 4, 4, 4] & [ 16, 16, 16] & [ 5.6, 5.6,11.0] & [ 3, 3, 1] & 6.5 & 4.0 & [   6.3,   6.3,   7.4] \\ 
199  &  $\text{Ba}_{4}\text{N}_{4}\text{Zn}_{2}$-139 & [ 7.1, 7.1, 7.1] & [ 4, 4, 4] & [ 14, 14, 14] & [ 4.2, 4.2,13.0] & [ 4, 4, 1] & 6.4 & 4.0 & [   2.3,   2.3,   3.0] \\ 
200  &  $\text{Mg}_{2}\text{Na}_{2}\text{Sb}_{2}$-129 & [ 4.6, 4.6, 7.7] & [ 6, 6, 4] & [ 22, 22, 14] & [ 4.6, 4.6, 7.7] & [ 3, 3, 2] & 6.5 & 4.0 & [   2.9,   2.9,   1.3] \\ 
201  &  $\text{Bi}_{2}\text{K}_{2}\text{Mg}_{2}$-129 & [ 4.9, 4.9, 8.5] & [ 6, 6, 4] & [ 20, 20, 12] & [ 4.9, 4.9, 8.5] & [ 3, 3, 2] & 6.5 & 4.0 & [   3.0,   3.0,   0.8] \\ 
202  &  $\text{Mg}_{4}\text{Se}\text{Te}_{3}$-115 & [ 4.5, 4.5,12.7] & [ 7, 7, 2] & [ 22, 22,  8] & [ 4.5, 4.5,12.7] & [ 3, 3, 1] & 6.0 & 4.0 & [  10.0,   9.8,   6.7] \\ 
203  &  $\text{As}_{4}\text{Sr}_{2}\text{Zn}_{4}$-164 & [ 4.3, 4.3, 7.3] & [ 7, 7, 4] & [ 24, 24, 14] & [ 7.4, 4.3, 7.3] & [ 2, 4, 2] & 6.5 & 3.0 & [   3.7,   3.6,   2.5] \\ 
204  &  $\text{Ga}_{4}\text{Sr}_{2}\text{Te}_{8}$-97 & [ 6.8, 6.8, 6.8] & [ 4, 4, 4] & [ 16, 16, 16] & [ 6.8, 8.4, 8.4] & [ 2, 2, 2] & 6.5 & 4.0 & [   2.3,   0.8,   0.7] \\ 
205  &  $\text{Ca}_{2}\text{Mg}_{4}\text{N}_{4}$-164 & [ 3.6, 3.6, 6.1] & [ 8, 8, 5] & [ 28, 28, 16] & [ 6.2, 3.6, 6.1] & [ 2, 4, 2] & 6.0 & 3.0 & [  11.8,  11.5,   9.4] \\ 
206  &  $\text{Cu}\text{Ga}_{5}\text{Te}_{8}$-111 & [ 6.0, 6.0,12.2] & [ 5, 5, 2] & [ 18, 18,  8] & [ 6.0, 6.0,12.2] & [ 3, 3, 1] & 6.5 & 4.0 & [   3.5,   3.5,   2.6] \\ 
207  &  $\text{Ca}_{4}\text{Si}_{4}\text{Sr}_{4}$-62 & [ 8.1, 4.9, 9.2] & [ 4, 6, 3] & [ 12, 20, 12] & [ 8.1, 4.9, 9.2] & [ 2, 3, 2] & 6.5 & 4.0 & [   1.9,   2.1,   2.0] \\ 
208  &  $\text{K}_{2}\text{Mg}_{2}\text{Sb}_{2}$-129 & [ 4.8, 4.8, 8.3] & [ 6, 6, 4] & [ 22, 22, 12] & [ 4.8, 4.8, 8.3] & [ 3, 3, 2] & 6.5 & 4.0 & [   4.7,   4.7,   1.4] \\ 
209  &  $\text{Cu}_{2}\text{S}_{4}\text{Sc}_{2}$-156 & [ 3.8, 3.8, 6.0] & [ 8, 8, 5] & [ 28, 28, 18] & [ 6.5, 3.8, 6.0] & [ 2, 4, 2] & 6.0 & 3.0 & [   3.0,   3.0,   2.2] \\ 
210  &  $\text{N}_{4}\text{Se}_{2}\text{Zr}_{4}$-63 & [ 3.7, 3.7,13.3] & [ 8, 8, 2] & [ 28, 28,  8] & [ 3.7, 3.7,13.3] & [ 4, 4, 1] & 6.5 & 4.0 & [  18.4,  18.4,   8.9] \\ 
211  &  $\text{Ba}_{3}\text{Na}_{3}\text{P}_{3}$-189 & [ 8.0, 4.7, 8.0] & [ 4, 6, 4] & [ 14, 22, 14] & [ 8.0, 4.7, 8.0] & [ 2, 3, 2] & 6.5 & 4.0 & [   1.2,   1.6,   1.1] \\ 
212  &  $\text{K}_{2}\text{Li}_{2}\text{Te}_{2}$-129 & [ 4.8, 4.8, 7.8] & [ 6, 6, 4] & [ 22, 22, 14] & [ 4.8, 4.8, 7.8] & [ 3, 3, 2] & 6.5 & 4.0 & [   2.8,   2.8,   1.5] \\ 
213  &  $\text{Bi}_{4}\text{Li}_{8}\text{Na}_{4}$-225 & [ 5.0, 5.0, 5.0] & [ 6, 6, 6] & [ 20, 20, 20] & [ 7.0, 7.0, 7.0] & [ 2, 2, 2] & 6.5 & 4.0 & [   1.8,   1.8,   1.8] \\ 
214  &  $\text{Ca}_{4}\text{N}_{4}\text{Zn}_{2}$-139 & [ 6.8, 6.8, 6.8] & [ 4, 4, 4] & [ 16, 16, 16] & [ 3.6, 3.6,12.6] & [ 4, 4, 1] & 6.3 & 3.0 & [   7.5,   7.5,  12.1] \\ 
215  &  $\text{Ca}_{3}\text{N}\text{Sb}$-221 & [ 4.9, 4.9, 4.9] & [ 6, 6, 6] & [ 22, 22, 22] & [ 4.9, 4.9, 4.9] & [ 3, 3, 3] & 6.5 & 4.0 & [   6.9,   6.9,   6.9] \\ 
216  &  $\text{As}_{4}\text{Se}_{4}\text{Y}_{4}$-62 & [ 3.9, 3.9,17.5] & [ 8, 8, 2] & [ 26, 26,  6] & [ 3.9, 3.9,17.5] & [ 4, 4, 1] & 5.0 & 3.0 & [  10.6,  10.6,   2.7] \\ 
217  &  $\text{Be}_{4}\text{Mg}_{2}\text{P}_{4}$-164 & [ 3.6, 3.6, 6.2] & [ 8, 8, 5] & [ 28, 28, 16] & [ 6.3, 3.6, 6.2] & [ 2, 4, 2] & 5.5 & 2.5 & [  14.6,  14.6,  11.0] \\ 
218  &  $\text{Cs}_{2}\text{Na}_{2}\text{S}_{2}$-129 & [ 4.8, 4.8, 8.0] & [ 6, 6, 4] & [ 22, 22, 14] & [ 4.8, 4.8, 8.0] & [ 3, 3, 2] & 6.5 & 4.0 & [   1.2,   1.2,   0.7] \\ 
219  &  $\text{N}_{4}\text{Sr}_{4}\text{Zn}_{2}$-139 & [ 7.0, 7.0, 7.0] & [ 4, 4, 4] & [ 14, 14, 14] & [ 3.9, 3.9,12.9] & [ 4, 4, 1] & 6.4 & 3.0 & [   5.4,   5.4,   7.7] \\ 
220  &  $\text{As}_{2}\text{K}_{2}\text{Li}_{4}$-59 & [ 4.5, 6.7, 6.4] & [ 7, 4, 5] & [ 22, 16, 16] & [ 4.5, 6.7, 6.4] & [ 3, 2, 2] & 6.0 & 4.0 & [   3.6,   2.0,   1.3] \\ 
221  &  $\text{Ba}_{2}\text{In}_{4}\text{Te}_{8}$-97 & [ 7.1, 7.1, 7.1] & [ 4, 4, 4] & [ 14, 14, 14] & [ 8.7, 8.7, 7.2] & [ 2, 2, 2] & 6.5 & 4.0 & [   0.7,   0.7,   2.5] \\ 
222  &  $\text{Ca}_{2}\text{N}_{4}\text{Zn}_{4}$-164 & [ 3.5, 3.5, 6.0] & [ 9, 9, 5] & [ 30, 30, 18] & [ 6.0, 3.5, 6.0] & [ 2, 4, 2] & 5.0 & 2.6 & [   7.5,   7.5,   6.1] \\ 
223  &  $\text{Li}_{2}\text{Na}_{2}\text{S}_{2}$-129 & [ 4.0, 4.0, 6.5] & [ 7, 7, 5] & [ 26, 26, 16] & [ 4.0, 4.0, 6.5] & [ 4, 4, 2] & 6.5 & 4.0 & [   3.9,   3.9,   3.3] \\ 
224  &  $\text{Cd}_{2}\text{In}_{4}\text{Te}_{8}$-82 & [ 7.8, 7.8, 7.8] & [ 4, 4, 4] & [ 14, 14, 14] & [ 6.4, 6.4,12.7] & [ 2, 2, 1] & 6.5 & 4.0 & [   2.5,   2.5,   3.3] \\ 
225  &  $\text{C}_{2}\text{Li}_{4}\text{N}_{4}$-139 & [ 5.1, 5.1, 5.1] & [ 6, 6, 6] & [ 20, 20, 20] & [ 3.7, 3.7, 8.7] & [ 4, 4, 2] & 6.5 & 3.0 & [   8.0,   8.0,   6.1] \\ 
\end{longtable*}
\end{center}

%

\end{document}